\documentclass[prb,showpacs,amsmath,amssymb,floatfix,superscriptaddress]{revtex4}

\usepackage{graphicx}
\usepackage{dcolumn}
\usepackage{bm}

\newcommand{\cf}[1]{\langle #1 \rangle}                      
\newcommand{\ket}[1]{\!\mid\! #1 \rangle}                    
\newcommand{\phdagger}{\mathop{\phantom{\dagger}}}           
\newcommand{\psiop}[1]{\psi^{\phdagger}_{#1}}                
\newcommand{\psidop}[1]{\psi^{\dagger}_{#1}}                 
\newcommand{\Psidop}[1]{\Psi^{\dagger}_{#1}}                 

\newcommand{\bml}{\begin{mathletters}}                           
\newcommand{\eml}{\end{mathletters} \hspace{-5pt}}       

\newcommand{\cbr}[1]{\left ( #1 \right )}

\begin{document}
\bibliographystyle{apsrev}

\title{Local Spectral Weight of a Luttinger Liquid: 
Effects from Edges and Impurities} 

\author{Paata Kakashvili}
\affiliation{Department of Applied Physics, Chalmers University of Technology,
SE 412 96 G\"oteborg, Sweden}
\author{Henrik Johannesson}
\affiliation{Department of Physics, G\"oteborg University, SE 412 96
G\"oteborg, Sweden}
\author{Sebastian Eggert}
\affiliation{Department of Applied Physics, Chalmers University of Technology,
SE 412 96 G\"oteborg, Sweden}
\affiliation{Department of Physics, University of Kaiserslautern,
D-67663 Kaiserslautern, Germany}

\begin{abstract}
We calculate the finite-temperature local spectral weight (LSW) of a Luttinger
liquid with an ''open'' (hard wall) boundary. Close to the boundary
the LSW exhibits characteristic oscillations indicative of spin-charge
separation. The line shape of the LSW is also found to have a
Fano-like asymmetry, a feature originating from the interplay between
electron-electron interaction and scattering off the boundary.
Our results can be used to predict how edges and impurities influence
scanning tunneling microscopy (STM) of one-dimensional electron systems
at low temperatures and voltage bias. Applications to STM on
single-walled carbon nanotubes are discussed.\\
\end{abstract}


\maketitle



\section{Introduction}
\label{section1}

Metallic electrons confined to one dimension exhibit a plethora of
intriguing effects, driven by interactions and the coupling to
impurities and defects \cite{Giamarchi}. At low energies a clean
system is described by the concept of a spinful {\em Luttinger liquid} (LL)
\cite{Voit}, with properties very different from those of a Fermi liquid: the 
quasiparticle pole vanishes identically and only collective modes
remain, separately carrying spin and charge. The response of an LL to
the addition of a local potential scatterer also differs dramatically
from that of a Fermi liquid: The repulsive electron-electron
interaction produces long-range density oscillations that get tangled
up with the impurity potential in such a way as to suppress the
single-electron spectral weight close to the impurity, as well as the
conductance through it \cite{KaneFisher,FurusakiNagaosa1}: In the
zero-temperature limit and with a spin-rotational invariant
interaction the impurity effectively cuts the system in two parts,
with ''open'' boundaries (hard walls) replacing the impurity.
The case of a magnetic impurity $-$ which interacts dynamically with the
conduction electrons $-$ is similar: In the zero-temperature limit
the physics is that of two LLs separated by open boundaries,
with the finite-$T$ response governed by a scaling operator that tunnels
electrons through the boundaries \cite{FurusakiNagaosa2}. The pictures that
emerge in both cases are universal in the sense that all response
functions depend only on the electron-electron interaction, with
critical exponents which for a spin-rotational interaction is coded by
the single LL {\em charge parameter} $K_c$. Details of the coupling
of the electrons to the impurity, or the structure of the impurity
potential, are irrelevant.

The fact that an impurity in an LL drives the system to an {\em open boundary
fixed point} \cite{EggertAffleck} has spurred considerable theoretical
work on properties of LLs with an open boundary condition (OBC)
\cite{FabrizioGogolin,EJM,MEJ,VoitWangGrioni,WangVoitPu,AnfusoEggert,Schonhammer1,Meden1}.
Added interest comes from the fact that many measurements on
one-dimensional electron structures $-$ such as the single-wall carbon
nanotubes (SWCNTs) \cite{Bockrath}, or quantum wires, realized in
gated semiconductor heterostructures \cite{Yacoby} or grown on
metallic substrates \cite{Segovia} $-$ are expected to be
significantly influenced by electron scattering from the edges, where
the confining potential to a first approximation can be treated as an OBC. 

Most work to date has focused on the {\em local spectral weight} (LSW)
of an LL with an OBC, yielding predictions for single-electron
tunneling and photoemission measurements close to an edge \cite{Eggert} 
or close to an impurity at sufficiently low temperatures \cite{Kivelson}. 
Measuring the energy $\omega$ (with $\hbar\!=\!1$) with respect to the
Fermi level, the low-temperature LSW $A(\omega)$ close to an open
boundary scales as \cite{KaneFisher,FabrizioGogolin,EJM}
\begin{equation}
\label{PowerLawBoundary}
A(\omega) \sim \omega^{(K_c^{-1}-1)/2}
\end{equation}
where $K_c < 1$ for a repulsive electron-electron interaction
\cite{CorrectionFootnote}. 
This is to be compared with that of a clean system probed away from its edges,
where $A(\omega) \sim \omega^{(K_c+K_c^{-1})/4 -1/2}$ \cite{Voit}.
Experiments on SWCNTs seem to agree with the theoretical prediction
that the tunneling rate of electrons should follow a characteristic
power law with temperature \cite{Bockrath}, with a significant
reduction of tunneling into the end of a tube as compared to tunneling
into its interior (''bulk'' regime) \cite{Yao}. Oscillation patterns
that suggest spin-charge separation have also been seen in the
tunneling conductance between two quantum wires produced by cleaved
edge overgrowth \cite{Auslaender}, in qualitative agreement with
theoretical results. In another line of research, photoemission
spectroscopy measurements on quasi-one-dimensional organic conductors
have been interpreted within a picture where the one-dimensional
chains in the samples are cut by impurities into disconnected pieces,
each modeled as an LL with OBCs. Again using results for the LSW, it
has been argued \cite{EJM,MEJ,VoitWangGrioni,Segovia,Zwick} that this
approach gives better agreement with experiments than conventional
theory where photoemission spectra are compared to predictions from
ordinary ``bulk'' LL theory \cite{Gunnarsson}. However, this
alternative interpretation remains controversial and the
issue has been difficult to settle, much due to the fact that photoemission
measurements on these materials are subject to a variety of subtle effects.

The most direct way to probe an LSW is via scanning tunneling microscopy (STM)
\cite{Eggert}. These experiments are delicate, as the STM tip must be
positioned at a very small distance from the sample for
electrons to tunnel \cite{SPM}. While this is feasible for SWCNTs, the
high-precision STM experiments that have been carried out have probed
tubes deposited on metallic substrates. This leads to a suppression of the 
electron-electron interaction from screening charges, and early results were
successfully interpreted within a free electron model \cite{Venema,Lemay}. 
In another effort STM measurements were performed on SWCNTs freely suspended 
over a trench \cite{LeRoy}, thus bypassing the problem with screening charges.
However, the resolution achieved in this experiment was not sufficient
to test for the expected LL scaling at small energies. 
In more recent experiments SWCNTs deposited on atomically clean
Au(111) surfaces were studied by high-resolution STM spectroscopy
\cite{Lee1}, revealing that the electronic standing waves close to the
end of a tube have en enhanced charge velocity which may imply 
spin-charge separation, and {\em a fortiori} LL behavior \cite{Lee}.

Turning to theory, the spectral properties of LLs with OBCs are by now
fairly well understood, although some open problems remain. Maybe most
pressing is the question about the very applicability of LL theory: What
is the energy scale $\Delta$ below which the power law in Eq.
(\ref{PowerLawBoundary}) becomes visible?
Obviously, an answer to this question is essential for making sensible
predictions for experiments. From numerical and other studies of the
one-dimensional Hubbard model \cite{Penc} it is known that the decrease
of the LSW $-$ as predicted by LL theory $-$ is often preceded by a
sharp increase, and that this effect is particularly pronounced near
an edge \cite{Schonhammer1,Meden1} or close to an impurity \cite{Meden2}.
The effect is expected to be generic for any one-dimensional
metallic system where the amplitude for back-scattering is larger than
for forward scattering. For some systems with a (weakly screened)
long-range interaction, such as the carbon nanotubes, back-scattering
gets suppressed above a threshold temperature, and one
expects the asymptotic LL scaling in Eq. (\ref{PowerLawBoundary})
to be visible at accessible energy scales, as is also suggested by experiments
\cite{Bockrath}. More work is needed, though, to obtain a reliable
estimate of the crossover scale $\Delta$, given data from
the underlying microscopic physics. 

We shall not address this issue here, but rather revisit the problem
of determining the full coordinate- and temperature dependence of the
LSW of one-dimensional interacting electrons with an OBC, 
assuming that the energy scale is sufficiently low for LL theory to
be applicable. Knowing the detailed structure of the LSW is important
for making predictions of future high-precision STM measurements of LL
systems, of which the SWCNTs are presently the prime candidates
\cite{EggerGogolinEPJ}. In earlier works the zero-temperature
properties of the LSW \cite{Eggert}, as well as the finite
temperature properties of the uniform part of the LSW (neglecting
Friedel oscillations) \cite{MEJ}, have been reported. 
Here we treat the full problem at a finite temperature and exhibit the
LSW for different choices of interaction strength and band filling. We
shall find that close to an open boundary the line shape of the LSW
has a marked asymmetry as a function of energy with respect to the
Fermi level, a property that arises from the phases that appear in the
single-electron Green's function, and which has not been examined
before. The form of the asymmetry in the neighborhood of the Fermi
level resembles a Fano line shape, a feature expected universally
whenever a resonant state (like that induced by a magnetic impurity in an 
electron system) interferes with a non-resonant one \cite{Fano}. As we
shall see, the origin of the asymmetric line shape in the present case
is very different, and is formed by an interplay between
electron-electron interaction and scattering off the open boundary.
The asymmetry is fairly robust against thermal effects, suggesting
that Fano-like line shapes produced by the reflection of interacting
electrons off boundaries can be observed at temperatures higher than
those originating from their interference with a resonating
level. Having access to the full LSW we will also be able to give a
systematic description of how charge- and spin separation shows up as an
oscillation pattern when being close to an edge (or, an impurity, at low
temperatures). This information, which we extract for different
temperatures, can be directly translated into a prediction of the measured
differential conductance when probing an LL system by STM. Also, given
the full LSW we derive its crossover from boundary to thermal scaling near the
Fermi level. The thermal effects soften the power law singularities of
the LSW, since the non-chiral terms in the electron Greens's function
produce a leading scaling term that is linear in energy at higher
temperatures.  This softening should not be confused with the
averaging effects that always occur when the experimental tunneling
currents are calculated by integrating over the Fermi-Dirac distribution. 

Our paper is organized as follows: In Sec.~II we review some basics about 
temperature-dependent local spectral weights and STM currents. In
Sec.~III we derive an exact representation of the local
spectral weight for a Luttinger liquid with an open boundary, paying
due attention to the phase dependence that has not been examined in
earlier studies. In this section we also show how to adapt
the theory for applications to scanning tunneling microscopy of
SWCNTs. Sec.~IV contains our results, and in Sec.~V we summarize the
most important points. A reader mostly interested in the physics of
the problem is advised to go directly to Sec.~IV. 
Unless otherwise stated we use units where $\hbar=k_B=c=1$.

\section{Preliminaries}
\label{section2}

In order to calculate tunneling currents e.g.~from an STM tip we will
consider the transition rate of adding electrons to an LL system
at a position $x$ and with energy $\omega$, 
\begin{equation}  \label{Overlap}
\Gamma^+(\omega, x;\beta) =  2 \pi g^2 Z^{-1}  \sum_{m, n} \exp(-\beta E_m) \,|
\cf{n | \Psidop{\sigma}(x) | m} |^2\,
\delta(\omega - E_n + E_m).
\end{equation}
This expression
follows from Fermi's golden rule assuming a tunneling Hamiltonian of the form 
$-g \Psidop{\sigma} \psi_{\sigma, \rm tip} + h.c.$, and treating the tip
as a reservoir with unit probability that an electron is available for
tunneling. Here $Z$ is the partition function of the $N$-particle system,
$\Psidop{\sigma}(x)$ creates an electron in the sample with spin
$\sigma$ at $x$, and $\psi^{\phantom{\dagger}}_{\sigma, \rm tip}$
removes an electron of the same spin from the tip. Equation (\ref{Overlap}) represents the probability that the $N$-particle states $\ket{m}$ of
energy $E_m$ are connected to the $(N+1)$-particle states  $\ket{n}$
of energy $E_n = E_m + \omega$ by the addition of an extra electron of
energy $\omega$ and coordinate $x$.  The transition rate $\Gamma^-$ of
removing an electron is given by Eq.~(\ref{Overlap}) by simply
replacing the index $m$ by $n$ in the Boltzmann weight,  assuming that
a ''hole'' is available in the tip with unit probability.

In order to calculate the transition rates it is useful to define the 
single-electron local spectral weight (LSW) $A(\omega,x;\beta)$ 
which is directly related to the transition rates by 
\begin{eqnarray}  \label{Aomega}
A(\omega, x;\beta) & = &  Z^{-1} (1+\exp(-\beta \omega)) \sum_{m, n} \exp(-\beta E_m) \,|
\cf{n | \Psidop{\sigma}(x) | m} |^2\,
\delta(\omega - E_n + E_m) \nonumber \\
& = & (1+\exp(-\beta \omega)) \Gamma^+(\omega, x;\beta)/2 \pi g^2 \\
& = & (1+\exp(\beta \omega)) \Gamma^-(\omega, x;\beta)/2 \pi g^2. \nonumber
\end{eqnarray}
It is well-known that the LSW defined in this way can be 
extracted from the spectral representation
of the single-electron retarded Green's function 
\begin{equation} \label{DEF_RET}
G^{R}(t,x,\beta) = -i \Theta(t) \cf{\{ \Psi_{\sigma}(t,x), \Psi^{\dag}_{\sigma}(0,x)
\}}_{\beta},
\end{equation}
by using that \cite{Rickayzen}
\begin{equation}  \label{LSW}
A(\omega,x;\beta) = -\frac{1}{\pi} \mbox{Im} \int_0^{\infty} G^{R}(t,x;\beta)
\mbox{e}^{i\omega t} dt.
\end{equation}
At zero temperature this quantity is known to be 
the single-electron local density of states $N(\omega,x)$ in agreement
with the definition in Eq.~(\ref{Aomega}).

We shall extract the LSW in the standard way by first calculating the 
single-electron retarded Green's function.
The calculation of $A(\omega,x;\beta)$ for the present problem
requires some care in order to analyze the analytic structure and
phase dependence in detail. In fact, the result which we derive in
the next section, using bosonization, reveals a surprising 
asymmetric energy dependence of the LSW close to the Fermi level
for a semi-infinite LL with an open boundary condition (OBC). 

Before taking on this task, let us recall how scanning tunneling
microscopy (STM) is used to experimentally probe the LSW close to
edges and impurities. In the simplest approach, when the STM tip is
assumed to couple only to the conduction electrons (thus neglecting
tunneling into localized impurity levels) the tunneling current
is given by the integrated difference between the transition rates
$\Gamma^{+}$ and $\Gamma^{-}$, weighted by the corresponding probabilities that an electron [hole] is available in the tip for tunneling 
to [from] the sample. With an applied voltage $V$ one thus has: 
\begin{eqnarray} \label{IT}
I(V,x;\beta) & = &  e
\int_{-\infty}^{\infty} N_{STM}(\omega-eV) \left[f(\omega-eV) 
\Gamma^+(\omega, x;\beta) - (1-f(\omega-eV)) \Gamma^-(\omega, x;\beta)\right]  d \omega
\nonumber \\
& \approx & 2 \pi e g^2 \rho_0 \int_{-\infty}^{\infty}
[f(\omega-eV)-f(\omega)]
 A(\omega, x; \beta) d \omega,
\end{eqnarray}
where $f(\omega)$ is the Fermi-Dirac distribution and 
we have approximated the density of states $N_{STM}(\omega)$ in the
tip by a constant $\rho_0$ in the last step.  
It is clear that we recover the conventional formula for
tunneling at zero temperature \cite{Tersoff}
\begin{equation} \label{I}
I(V,x) = {2e \pi g^2} \int_{0}^{eV} N(\omega, x) N_{STM}(\omega-eV) d \omega,
\end{equation}
where $N(\omega, x)$ is the local single-electron density of states
for a conduction electron in the sample, and $N_{STM}(\omega)$ is the
density of states of the STM tip measured relative to the Fermi energy.  
By differentiating, the local differential tunneling conductance can then be 
directly related to the local density of states in Eq. (\ref{I})
\begin{equation} \label{LDOS}
\frac{dI(V,x)}{dV} \propto 2e^{2} \pi g^2 \rho_0 N(V,x).
\end{equation}
This expression remains valid at a finite temperature $T$, provided
that the thermal length $\lambda_T \sim v_s/T$
is larger than any other characteristic length $L$ of the experimental
setup (such as the distance between the STM tip and the edge of the sample).
The speed $v_s$ that determines $\lambda_T$ is that
of the spin collective modes (which in a one-dimensional interacting
electron system are slower than the collective charge modes). When $L
> \lambda_T$ a temperature-dependent description becomes necessary,
and the expression for $I(V,x)$ has to be  modified according to
Eq.~(\ref{IT}). The local differential conductance at finite
temperature can therefore be written as
\begin{equation}\label{LDOST}
\frac{dI(V,x;\beta)}{dV} = 2e g^2 \pi \rho_0 \int_{-\infty}^{\infty}
\frac{d}{dV}f(\omega-eV)
 A(\omega, x;\beta) d \omega.
\end{equation}
It follows that the line shape properties of the local tunneling
conductance are directly determined by the LSW.

With these preliminaries we now turn to the calculation of the
finite-temperature LSW for an LL with an open boundary.

\section{Deriving the local spectral weight}
\label{section3}

We consider an interacting electron liquid on a semi-infinite line, $x \ge 0$,
subject to an OBC at the end $x=0$.
Following standard Luttinger-liquid approach \cite{Giamarchi}, we linearize the
spectrum and 
decompose the electron field $\Psi_{\sigma}$
into left- $(L)$ and right- $(R)$ moving chiral fermions at the two
Fermi points $\pm k_F$,
\begin{equation} \label{Decomp}
\Psi_{\sigma}(x)=e^{-ik_{F}x} \psiop{L\sigma}(x)
+e^{ik_{F}x} \psiop{R\sigma}(x).
\end{equation}
The zero-temperature single-electron Green's function at a point $x$
can then be expressed in terms of the propagators of the time-evolved
chiral fermions
\begin{multline}\label{GFNC}
G(t>0,x)=\cf{\Psi_{\sigma}(t,x) \Psi_{\sigma}^{\dag}(0,x)}=
\cf{\psiop{L\sigma}(t,x) \psidop{L\sigma}(0,x)}+
\cf{\psiop{R\sigma}(t,x) \psidop{R\sigma}(0,x)}\\ 
+e^{i2k_{F}x}\cf{\psiop{R\sigma}(t,x) \psidop{L\sigma}(0,x)}+e^{-i2k_{F}x}
\cf{\psiop{L\sigma}(t,x) \psidop{R\sigma}(0,x)}. 
\end{multline}
We see that there are two types of contributions to $G(t\!>0,x)$:
oscillatory and non-oscillatory. While the latter are always present,
the former are nonzero only if the left- and right moving fermions get
entangled at a boundary. Imposing an open {\em (Dirichlet)} boundary
condition at the ''phantom site'' which is situated one lattice
spacing $a$ from the end of the LL at $x=-a$,
\begin{equation}
\Psi_{\sigma}(-a)=e^{ik_{F}a}\psiop{L\sigma}(-a)
+e^{-ik_{F}a}\psiop{R\sigma}(-a)=0,
\end{equation}
and assuming that the chiral fermions are slowly varying on the scale
of $a$, it follows that
\begin{equation}\label{BC}
\psiop{R\sigma}(0)=e^{i\gamma}\psiop{L\sigma}(0),
\end{equation}
where
\begin{equation}
\gamma=\pi+2k_{F}a=\pi(1+n_{e}), \label{gamma}
\end{equation}
 with $n_e$ the filling factor ($n_e\!=\!1$ for
a half-filled band). Although not essential here, the "softening" of
the boundary $-$ implied by imposing the Dirichlet condition at
$x\!=\!-a$\, $-$ is sometimes useful for modeling the dependence of
the scattering phase shift $\gamma$ on the shape of the edge- or
impurity-potential.  The value of $\gamma$ may therefore depend on the
details of the boundary geometry, but it is important to notice that
it is in general not a multiple of $\pi$ even at half-filling. Using
Eq. (\ref{BC}) to analytically continue to negative coordinates
\cite{EggertAffleck}, the right-movers may be represented by left-movers as
\begin{equation} \label{AnalyticCont}
\psiop{R\sigma}(x)=e^{i\gamma}\psiop{L\sigma}(-x), \ \ x > 0.
\end{equation}
We can then express the Green's function in Eq. (\ref{GFNC}) in terms
of left-moving fermions only, now taking values on the full line
$-\infty <x<\infty$
\begin{multline}\label{GFC}
G(t>0,x)=\cf{\psiop{L\sigma}(t,x) \psidop{L\sigma}(0,x)}+
\cf{\psiop{L\sigma}(t,-x) \psidop{L\sigma}(0,-x)} \\
+e^{i(2k_{F}x+\gamma)}\cf{\psiop{L\sigma}(t,-x)\psidop{L\sigma}(0,x)}+
e^{-i(2k_{F}x+\gamma)} \cf{\psiop{L\sigma}(t,x) \psidop{L\sigma}(0,-x)}.
\end{multline}
Introducing 
\begin{equation} \label{Chiral}
G_{LL}(t,x,x')=\cf{\psiop{L\sigma}(t,x) \psidop{L\sigma}(0,x')}
=
\cf{\psidop{L\sigma}(t,x) \psiop{L\sigma}(0,x')},
\end{equation}
with the second equality following from the charge conjugation symmetry of the
linearized theory, Eq. (\ref{GFC}) may be written as
\begin{equation} \label{GFCrewritten}
G(t>0,x)=G_{LL}(t,x,x)+G_{LL}(t,-x,-x) +e^{i(2k_{F}x+\gamma)}G_{LL}(t,-x,x)
+e^{-i(2k_{F}x+\gamma)}G_{LL}(t,x,-x).
\end{equation}
With the definition in Eq. (\ref{DEF_RET}) the 
retarded Green's function can finally be cast on the compact form
\begin{equation}\label{RGFC}
G^{R}(t,x)=-i \Theta(t)(4ReG_{LL}(t,x,x)+2e^{i(2k_{F}x+\gamma)}ReG_{LL}(t,-x,x) \\
+2e^{-i(2k_{F}x+\gamma)}ReG_{LL}(t,x,-x)),
\end{equation}
using Eqs. (\ref{Chiral}) and (\ref{GFCrewritten}).
To obtain the LSW in Eq. (\ref{LSW}) we thus need to calculate the chiral
Green's function in (\ref{Chiral}), identify its real part, and then Fourier
transform the resulting expression for $G^{R}(t,x)$ from (\ref{RGFC}). 
The first part can be done analytically
by using bosonization, and we turn to this task in the next section.

\subsection{Chiral Green's function from bosonization}

Using standard bosonization \cite{GNT} we write
the left- and right-moving fermion fields as coherent superpositions
of free bosonic charge and spin fields, $\phi_{rc} = (\phi_{r\uparrow}
+ \phi_{r\downarrow})/\sqrt{2}$ and $\phi_{rs}= (\phi_{r\uparrow} -
\phi_{r\downarrow})/\sqrt{2}$, with $r=L, R$
\begin{eqnarray} \label{Bosonization}
\psiop{L\sigma}(t,x)&\!=\!&\frac{\eta_{L\sigma}}{\sqrt{2\pi \alpha}}
\exp \left(- i \sqrt{2\pi} (\cosh\theta \,\phi_{Lc}(x,t)
+\sinh\theta \,\phi_{Rc}(x,t) + \sigma \phi_{Ls}(x,t) )\right)
\nonumber \\
\psiop{R\sigma}(t,x)&\!=\!&\frac{\eta_{R\sigma}}{\sqrt{2\pi \alpha}}
\exp \left( i \sqrt{2\pi} (\cosh\theta \,\phi_{Rc}(x,t)
+\sinh\theta \,\phi_{Lc}(x,t) + \sigma \phi_{Rs}(x,t) )\right).
\end{eqnarray}
Here $\alpha$ is a small-distance cutoff of the order of the lattice
spacing of the underlying microscopic model, and $\eta_{r\sigma}$ are
Klein factors obeying a diagonal Clifford algebra that ensure that
fermion fields of different chirality $r$ and/or spin $\sigma$ anticommute.  
The parameter $\theta$ is related to the LL charge parameter $K_c$ by
$K_c = \mbox{e}^{2\theta}$, and is parameterized by the amplitudes for the
low-energy scattering processes.
%
%
%
%
For a system with long-range interaction these amplitudes become
momentum dependent, but since we shall only be interested in the
asymptotic low-energy behavior of the Green's functions we can restrict them 
to zero momentum and treat $K_c$ as a constant (taking a value $<\!1$ for
repulsive interaction). Note that this shortcut assumes a finite-range
interaction, whereas an unscreened Coulomb interaction which diverges
at vanishing momentum leads to very different physics \cite{Schulz}. 
Away from a half filling umklapp scattering vanishes, and standard RG
arguments show that backscattering processes become irrelevant (again
assuming a repulsive electron-electron interaction). 
As is well-known, for this case the remaining 
dispersive and forward scattering vertices can be written as quadratic forms in
bosonic operators, leading to two free boson theories, one for charge,
and one for spin
\begin{equation} \label{BosonHamiltonian}
{\cal H} = 
\sum_{j=c,s}\frac{v_{j}}{2} \left( (\partial_x \phi_{Lj})^2 + (\partial_x
\phi_{Rj})^2 \right).
\end{equation}
This defines the LL Hamiltonian density, here expressed in the chiral fields,  
with $v_{c\,(s)}$ the speed of the charge (spin) bosonic modes.

The logic of the construction just sketched is strictly valid only for a
translational invariant system where all interaction processes can be
classified into dispersive, forward, backward, or umklapp scattering
\cite{Giamarchi}. For a system with an open boundary, translational
invariance is broken and a two-particle interaction leads to
additional scattering processes. As shown by Meden {\em et al.}
\cite{Meden1}, however, the theory in Eq. (\ref{BosonHamiltonian})
still captures the universal low-energy physics. Perturbative
arguments suggest that the energy range where it applies increases 
with the range of the interaction of the original microscopic theory. 

To make progress we analytically continue the charge $\phi_{Lc}$ and
spin $\phi_{Ls}$ boson fields in (\ref{Bosonization}) to $x<0$ such
that the boundary condition in Eq.~(\ref{BC}) is satisfied
\begin{eqnarray} \label{AnalytCont2}
\phi_{Lc}(t,-x)&=&-\phi_{Rc}(t,x) + \frac{\gamma}{\sqrt{2\pi K_c}}
\nonumber \\
\phi_{Ls}(t,-x) &=&-\phi_{Rs}(t,x),  \ \ \ x>0.
\end{eqnarray}
Using (\ref{AnalytCont2})  we can write
\begin{equation} \label{ChiralField2}
\psiop{L\sigma}(t,x)=\frac{e^{i\gamma (1-K_c)/2K_c} \eta_{L\sigma}}
{\sqrt{2\pi \alpha}} 
\times \exp \left(- i \sqrt{2\pi} (\cosh\theta \,\phi_{Lc}(t,x)
-\sinh\theta \,\phi_{Lc}(t,-x) + \sigma \phi_{Ls}(t,x) )\right).
\end{equation}
Given (\ref{ChiralField2}) the chiral Green's functions in
(\ref{RGFC}) are now easily calculated, using that $\phi_{Lc/s}$ are
chiral bosons governed by the free theory in
(\ref{BosonHamiltonian}) \cite{GNT}
\begin{eqnarray} \label{LL1}
G_{LL}(t,x,x)&\!=\!&\left[ \frac{1}{\alpha+iv_{s}t} \right]^{1/2}\!
\left[ \frac{1}{\alpha+iv_{c}t} \right]^{k_{1}+k_{2}}\!
\left[ \frac{4x^{2}}{(\alpha+iv_{c}t)^{2}+4x^{2}} \right]^{k_{3}}, \\
\label{LL2}
G_{LL}(t,x,-x)&\!=\!&\left[ \frac{1}{\alpha\!+\!i(v_{s}t+2x)} \right]^{1/2}\!
\left[\frac{1}{\alpha\!+\!i(v_{c}t+2x)} \right]^{k_{1}}\!
\left[\frac{1}{\alpha\!+\!i(v_{c}t-2x)} \right]^{k_{2}}\!
\left[\frac{4x^{2}}{(\alpha\!+\!iv_{c}t)^{2}} \right]^{k_{3}},
\end{eqnarray}
with $G_{LL}(t,-x,-x)$ and $G_{LL}(t,-x,x)$ obtained from (\ref{LL1}) and
(\ref{LL2}) respectively by taking $x \rightarrow -x$. The exponents
are given by
\begin{eqnarray} \label{exponents}
k_{1}&=&\cbr{K_c+1/K_c+2}/8 \nonumber \\
k_{2}&=&\cbr{K_c+1/K_c-2}/8  \\
k_{3}&=&\cbr{1/K_c-K_c}/8 \nonumber
\end{eqnarray}

To obtain the finite-temperature Green's function we use conformal
field theory techniques \cite{DiFransesco} and map the complex planes
$\{z_j=v_j\tau+ix\}$ on which the zero-temperature chiral theory is defined
(with $\tau\!=\!it$ the Euclidean time, and $j=c,s$) onto two infinite
cylinders $\Gamma_j=\{w = v_j\tau'+ix'\}$ of circumference $\beta = 1/T$
\begin{equation}
w_j=\frac{v_j \beta}{\pi}
\arctan \left ( \frac{\pi}{\beta v_j}z_j \right ).
\end{equation}
Employing the transformation rule  
\begin{equation} \label{CFTtransformation}
\langle e^{i \alpha_1 \phi(w_1)} ... e^{i \alpha_n \phi(w_n)}\rangle=
\prod_{i=1}^{n} \left ( \frac{dw}{dz}\right)^{-\alpha_{i}^{2}/8 \pi}_{w=w_i}
\langle e^{i \alpha_1 \phi(z_1)} ... e^{i \alpha_n \phi(z_n)}\rangle
\end{equation}
we obtain for the finite-temperature versions of the Green's functions
\begin{multline} \label{GLL1}
G_{LL}(t,x,x;\beta)=\left(\frac{\pi}{\beta v_{c}}\right)^{k_{1}+k_{2}}
\left(\frac{\pi}{\beta v_{s}}\right)^{1/2}
\left[ \frac{1}{\sin(\frac{\pi}{\beta}(\alpha+it))} \right]^{1/2}
\left[ \frac{1}{\sin(\frac{\pi}{\beta}(\alpha+it))} \right]^{k_{1}+k_{2}}
\\
\times\left[ \frac{\sinh^{2}(\frac{\pi}{\beta}\frac{2x}{v_{c}})}
{\sin(\frac{\pi}{\beta} (\alpha+i(t+2x/v_{c})))
\sin (\frac{\pi}{\beta}(\alpha+i(t-2x/v_{c})))} \right]^{k_{3}}\!\!\!\!\!,
\end{multline}

\begin{multline} \label{GLL2}
\!\!\!\!\!\!G_{LL}(t,x,-x;\beta)=\left(\frac{\pi}{\beta v_{c}}\right)^{k_{1}+k_{2}}
\left(\frac{\pi}{\beta v_{s}}\right)^{1/2}
\!\!\left[ \frac{1}{\sin(\frac{\pi}{\beta}(\alpha+i(t+2x/v_{s})))}
  \right]^{1/2} \!
\left[ \frac{1}{\sin(\frac{\pi}{\beta}(\alpha+i(t+2x/v_{c})))}
  \right]^{k_{1}} 
 \\
\times \left[ \frac{1}{\sin(\frac{\pi}{\beta}(\alpha+i(t-2x/v_{c})))}
  \right]^{k_{2}} 
\left[ \frac{\sinh^{2}(\frac{\pi}{\beta}\frac{2x}{v_{c}})}
{\sin^{2}(\frac{\pi}{\beta}(\alpha+it))} \right]^{k_{3}}\!\!\!\!\!.
\end{multline}
We have here dropped the primes on the transformed 
coordinates and reinserted the real time variable. 

\subsection{An exact representation of the LSW}
\begin{figure}[!htb]
\begin{center}
\includegraphics[width=3.2in]{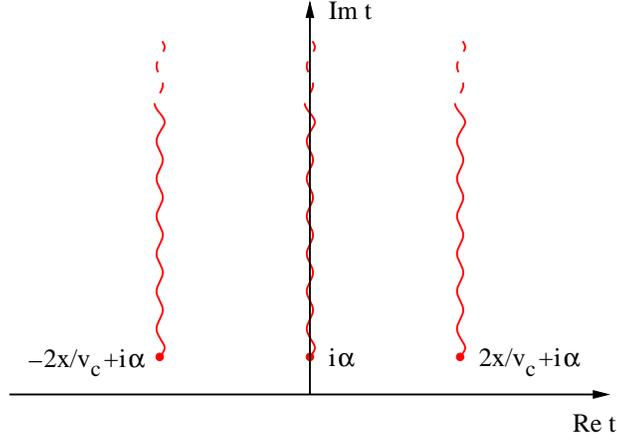}
\caption{[color online] Branch points and branch cuts for
  $G_{LL}(t,x,x;\beta)$. The phases are different in different
  regions defined by branch points.}
\label{branch_cut}
\end{center}
\end{figure}

Having obtained the chiral Green's functions we next calculate their
real parts, which, according to Eq. (\ref{RGFC}), define the LSW. It is
essential here to consistently identify the phases of the Green's functions
that appear in (\ref{RGFC}). This is most easily done by exponentiating
the expressions in (\ref{GLL1}) and (\ref{GLL2}) and choosing the negative
real axis as branch cut of the logarithm \cite{Mahan}. Treating $t$ as a
complex variable, this amounts to choosing
the branch cuts for $G_{LL}(t,x,x;\beta)$ as shown on Fig.~\ref{branch_cut}.
Functions $G_{LL}(t,x,x;\beta)$ and $G_{LL}(t,x,-x;\beta)$ have different 
phases in different regions defined by branch points, since the phases
differ by $2\pi$ on opposite sides of the cuts. Taking the cutoff
$\alpha \rightarrow 0$ we get
\begin{equation} 
{\rm Re}\,G_{LL}(t,x,x;\beta)=\left(\frac{\pi}{\beta v_{c}}\right)^{k_{1}+k_{2}}\!\!\!
\left(\frac{\pi}{\beta v_{s}}\right)^{1/2}\!\!\! \cos \zeta (t) 
\left |\sinh(\frac{\pi}{\beta}t) \right |^{-(1/2+k_1+k_2)}
\left| \frac{\sinh(\frac{\pi}{\beta} (t+\frac{2x}{v_{c}}))
\sinh(\frac{\pi}{\beta}(t-\frac{2x}{v_{c}}))}{\sinh^{2}(\frac{\pi}{\beta}\frac{2x}
{v_{c}})}  
 \right|^{-k_{3}}\!\!\!\!\!\!\!\!, 
\end{equation}
\begin{multline}
\label{RealParts}
{\rm Re}\,G_{LL}(t,x,-x;\beta)= \left(\frac{\pi}{\beta v_{c}}\right)^{k_{1}+k_{2}}\!\!\!
\left(\frac{\pi}{\beta v_{s}}\right)^{1/2}\!\!\! \cos \zeta' (t) 
\left |\sinh(\frac{\pi}{\beta} (t+2x/v_{s})) \right |^{-1/2} \\
 \times \left |\sinh(\frac{\pi}{\beta} (t+2x/v_{c})) \right |^{-k_{1}} 
\left |\sinh(\frac{\pi}{\beta} (t-2x/v_{c})) \right |^{-k_{2}}
\left|
\frac{\sinh(\frac{\pi}{\beta}t)}{\sinh(\frac{\pi}{\beta}\frac{2x}
{v_{c}})} \right|^{-2k_{3}}\!\!\!\!\!\!\!\!\!\!\!,
\end{multline}
with $k_1, k_2,$ and $k_3$ defined in Eq. (\ref{exponents}), and
where
\begin{eqnarray} \label{Phases}
\zeta(t)&=&\left\{
\begin{array}{ll}
\frac{\pi}{4}+\frac{\pi k_{1}}{2}+\frac{\pi k_{2}}{2} & 0<t<\frac{2x}{v_{c}}\\
\frac{\pi}{4}+\frac{\pi k_{1}}{2}+\frac{\pi k_{2}}{2}+\pi k_{3} &
t>\frac{2x}{v_{c}} 
\end{array} \right.
\nonumber \\
\zeta'(t)&=&\left\{
\begin{array}{ll}
\frac{\pi}{4}+\frac{\pi k_{1}}{2}+\frac{\pi k_{2}}{2}+\pi k_{3} &
t<-\frac{2x}{v_{s}} \\
-\frac{\pi}{4}+\frac{\pi k_{1}}{2}+\frac{\pi k_{2}}{2}+\pi k_{3} &
-\frac{2x}{v_{s}}<t<-\frac{2x}{v_{c}}\\
-\frac{\pi}{4}-\frac{\pi k_{1}}{2}+\frac{\pi k_{2}}{2}+\pi k_{3} &
-\frac{2x}{v_{c}}<t<0\\ 
-\frac{\pi}{4}-\frac{\pi k_{1}}{2}+\frac{\pi k_{2}}{2}-\pi k_{3} &
0<t<\frac{2x}{v_{c}}\\
-\frac{\pi}{4}-\frac{\pi k_{1}}{2}-\frac{\pi k_{2}}{2}-\pi k_{3} &
t>\frac{2x}{v_{c}} 
\end{array} \right.
\end{eqnarray}
We see that the phases that appear in the real parts of the Green's
functions take different values in different domains, consistent with
the fact that the end points of the domains are branch points (see
Fig.~\ref{branch_cut}).
Given the results in Eq. (\ref{RealParts})
we can finally write down an exact representation of the LSW for an LL
with an open boundary. Combining (\ref{LSW}) and (\ref{RGFC}) we find that
\begin{eqnarray} \label{LDOSF}
A(\omega,x,\beta)&=&\frac{4}{\pi}\int_{0}^{\infty} dt \cos \omega t
ReG_{LL}(t,x,x;\beta)
\nonumber \\
&+&\frac{2}{\pi}
  \int_{-\infty}^{\infty} dt \cos(2k_Fx + \gamma - \omega t) Re\,G_{LL}(t,x,-x;\beta).
\end{eqnarray}
This expression reveals that $A(\omega,x,\beta)$ has
a nontrivial dependence on the interaction strength (via $k_1, k_2,
k_3$ and the velocities $v_c$ and $v_s$ that parameterize the chiral
Green's functions), in addition to an oscillation in the second term
that is shifted by a phase controlled by the band filling via $k_F$
and $\gamma$ in Eq.~(\ref{gamma}) and the distance $x$ to the boundary. 
This leads to an interesting asymmetric dependence of the LSW on energy 
as will be discussed in Sec.~IV.

One of the most promising candidates for LL behavior are single-walled carbon
nanotubes (SWCNT), which however require some modification to 
the Green's functions in Eqs.~(\ref{GLL1})-(\ref{RealParts})
due to the presence of two electronic channels.
Depending on the substrate the interaction constant $K_c$ may be very
small $\approx 0.25$ and strongly k-dependent for isolating substrates
or closer to unity $\approx 0.6-1$ for metallic substrates.
Especially in the latter case backscattering processes may also play
an important role at very low temperatures.
Following [\onlinecite{EggerGogolinPRL,BalentsFisher}] the low energy
expression for the electron field is written in terms of Bloch
functions on the two graphite sublattices in order to arrive at two
spinful channels of 1D Fermion operators ${\psi}_{L \alpha \sigma}$
and ${\psi}_{R \alpha \sigma}$ where  $\sigma=\uparrow,\downarrow$ is
the spin and $\alpha=\pm$ labels the two distinct channels.
In order to bosonize the problem it is then possible to introduce
bosonic fields of total and relative $(\delta=\pm)$
charge and spin chiral bosons $\phi_{r j \delta} \, (r= L, R, \ \ j = c,s)$
and write the bosonization formula
\begin{equation} \label{SWNTBosonization}
{\psi}_{L \alpha \sigma} = \frac{\eta_{L \alpha \sigma}}{\sqrt{2\pi \alpha}}
\exp (-i\sqrt{\pi}[\cosh\theta\phi_{L c +}(x) + \sinh\theta \phi_{R c +}(x) 
+ \alpha \phi_{L c -}(x) + \sigma \phi_{L s +}(x) + \alpha \sigma \phi_{L s -}(x)]),
\end{equation}
where  ${\psi}_{R \alpha \sigma}$ is obtained by taking the complex
conjugate of the right-hand-side of Eq. (\ref{SWNTBosonization}) and
switching $L \leftrightarrow R$.

An open end or an impurity will in general mix the two channels so the
effective "open" boundary condition will in most cases not be a simple
reflection. The analytic continuation analogous to
Eq.~(\ref{AnalytCont2}) may therefore turn out to be more complicated
for the four bosons.  In the simplest symmetric cases we see that the
boundary Green's functions have again a structure as given in
Eqs.~(\ref{GLL1}-\ref{RealParts}), however with the exponent $1/2$
replaced by $3/4$ and 
$k_{1}=\cbr{K_{c+}+1/K_{c+}+2}/16$, $k_{2}=\cbr{K_{c+}+1/K_{c+}-2}/16$ and 
$k_{3}=\cbr{1/K_{c+}-K_{c+}}/16$.   Accordingly 
$\zeta (t)$ and $\zeta' (t)$ are given by Eq.~(\ref{Phases}) with each
occurrence of the term $\pi/4$ in (\ref{Phases}) replaced by $3\pi/8$.
The oscillating Friedel-like terms in the LSW in the second term of
Eq.~(\ref{LDOSF}) turn out to have the opposite sign on the two
sublattices \cite{mele}.

\section{Results}

In order to extract the physics from the LSWs derived in the previous
section we have numerically carried out the Fourier transforms in
Eq.~(\ref{LDOSF}). The results reveal a rich structure in the LSW of a
Luttinger liquid when an edge or an impurity is present.
We here focus on three aspects of the LSW of particular interest for
STM experiments: Its asymmetric line shape as a function of energy,
oscillation patterns revealing spin-charge separation, and thermal effects.
To keep the discussion general, we do not specify the dependence of the velocities
$v_{c,s}$ on the Luttinger liquid parameter $K_{c}$, which is sensitive to the
choice of microscopic model. Since our results do not depend on the
particular relation between $v_{c,s}$ and $K_{c}$, we choose them freely so as to
be able to observe all possible scenarios for the behavior of the LSW. The
question whether a particular scenario will be realized in a particular model is
of course determined by the relation between velocities and Luttinger liquid
parameter. 


\subsection{Properties of the LSW}

One of the most striking feature of the LSW of an LL is the well-known
power law suppression at low energies proportional to
$\omega^{(K_c+K_c^{-1})/4 -1/2}$. For the special case of zero temperature and 
commensurate oscillations (i.e.
setting $2k_F n a + \gamma = \pi m \, (m \in {\mathbb{Z}})$) 
the LSW of Eq.~(\ref{LDOSF}) near a boundary at $x=na$ has been 
analyzed before \cite{EJM,Eggert}.  In that case the first term of 
Eq.~(\ref{LDOSF}) takes on a scaling form with the variable $\omega x$ 
that shows a crossover from bulk scaling  to
boundary scaling in Eq.~(\ref{PowerLawBoundary}) with slow oscillations 
proportional to $\sin(2 \omega
x/v_c)\omega^{k_3-1}x^{k_3-k_1-k_2-1/2}$ \cite{EJM}.  
If the phase is neglected or set to $2 k_F n a + \gamma = m \pi$
the second Friedel-like term in Eq.~(\ref{LDOSF}) also takes on a
scaling form, but shows both spin and charge modulations that decay
with $\omega^{k_1-1}x^{-k_2-1/2}$.  
We will now focus on the energy dependence
at both small and large temperatures together with the effect of the
phase $\gamma$.
\begin{figure}[h!tb]
\begin{center}
\includegraphics[width=3.2in]{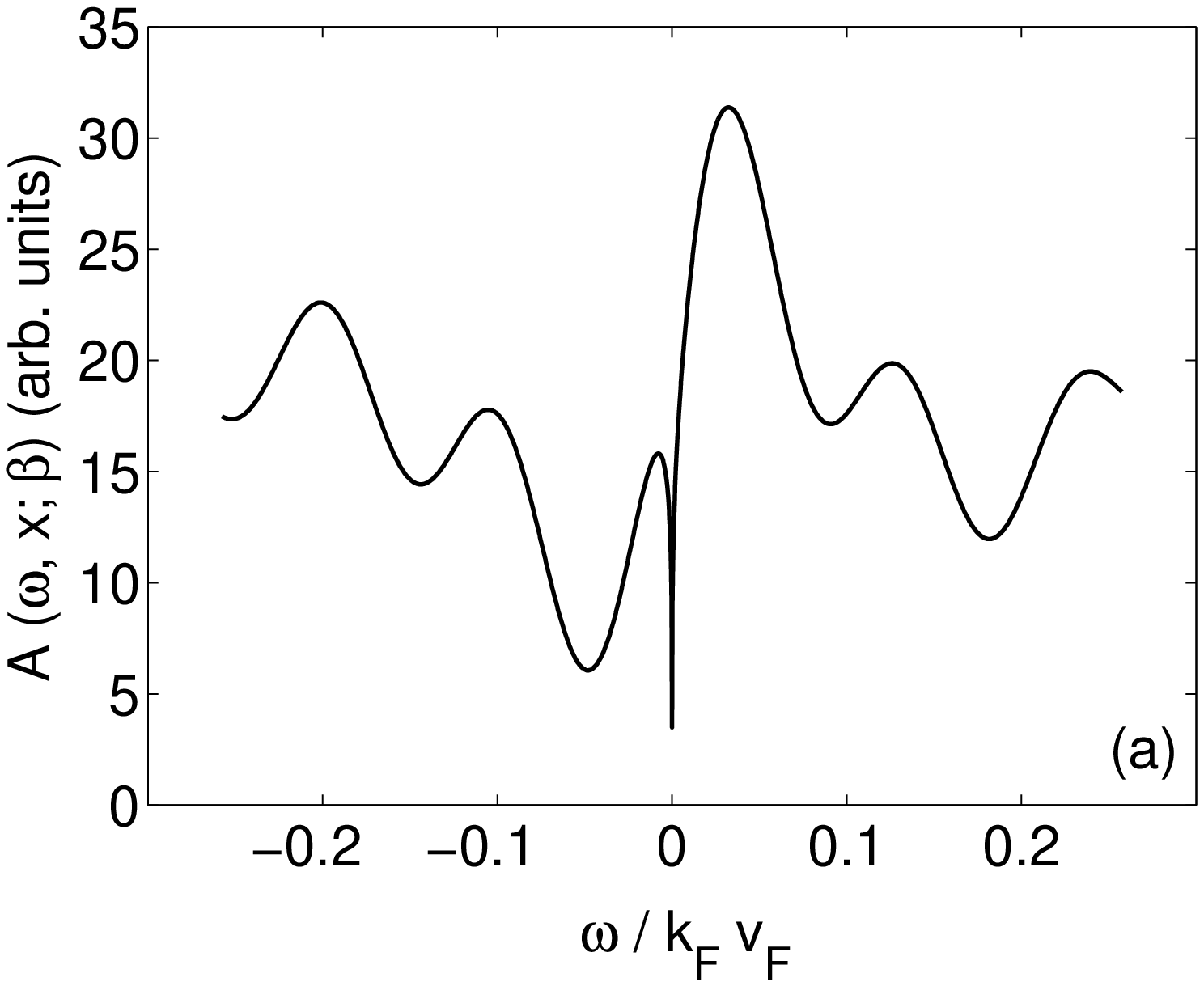}
\includegraphics[width=3.2in]{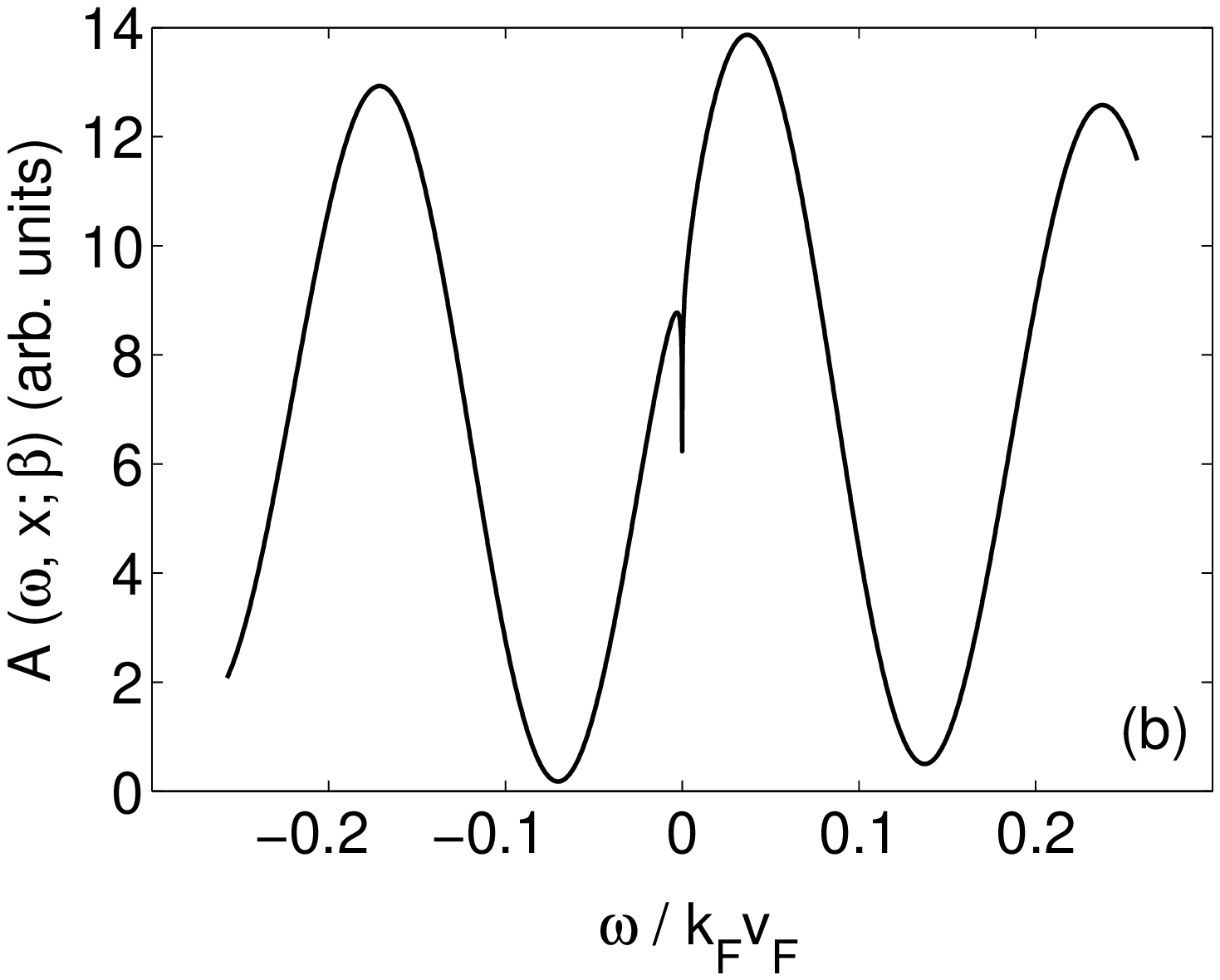}
\caption{Local spectral weight $A(\omega, x; \beta)$ for $K_{c}=0.7$,
  $v_{c}/v_{F} \approx 1.43$, $v_{c}/v_{s} \approx 3$ (a),
  $K_{c}=0.9$,  $v_{c}/v_{F} \approx 1.11$, $v_{c}/v_{s} \approx 1.26$ (b)
  with $T/k_{F}v_{F}=2.6 \times 10^{-6}$, $x=10a$, $n_{e}=0.97$.} 
\label{asymmetry}
\end{center}
\end{figure}

\subsubsection{Asymmetric line-shape} 
At small energies, the most striking feature of the LSW is the
asymmetry of its line shape, as seen in Fig.~\ref{asymmetry}. From
Eq.~(\ref{LDOSF}) it is apparent that this property is due to the
phase $2k_F x + \gamma$ appearing in the second term of $A(\omega, x;
\beta)$. Here the distance from the boundary $x=na$ has to be measured
in integer units of the lattice spacing $a$, since the original
Fermion operators typically correspond to orbitals in the crystal lattice.
The phase causes a shift of the periodic structure of $A(\omega, x;
\beta)$ with respect to the Fermi level, and accordingly determines
how the spectral weight suppression close to the Fermi level affects
the line shape.
\begin{figure}[h!bt]
\begin{center}
\includegraphics[width=3.2in]{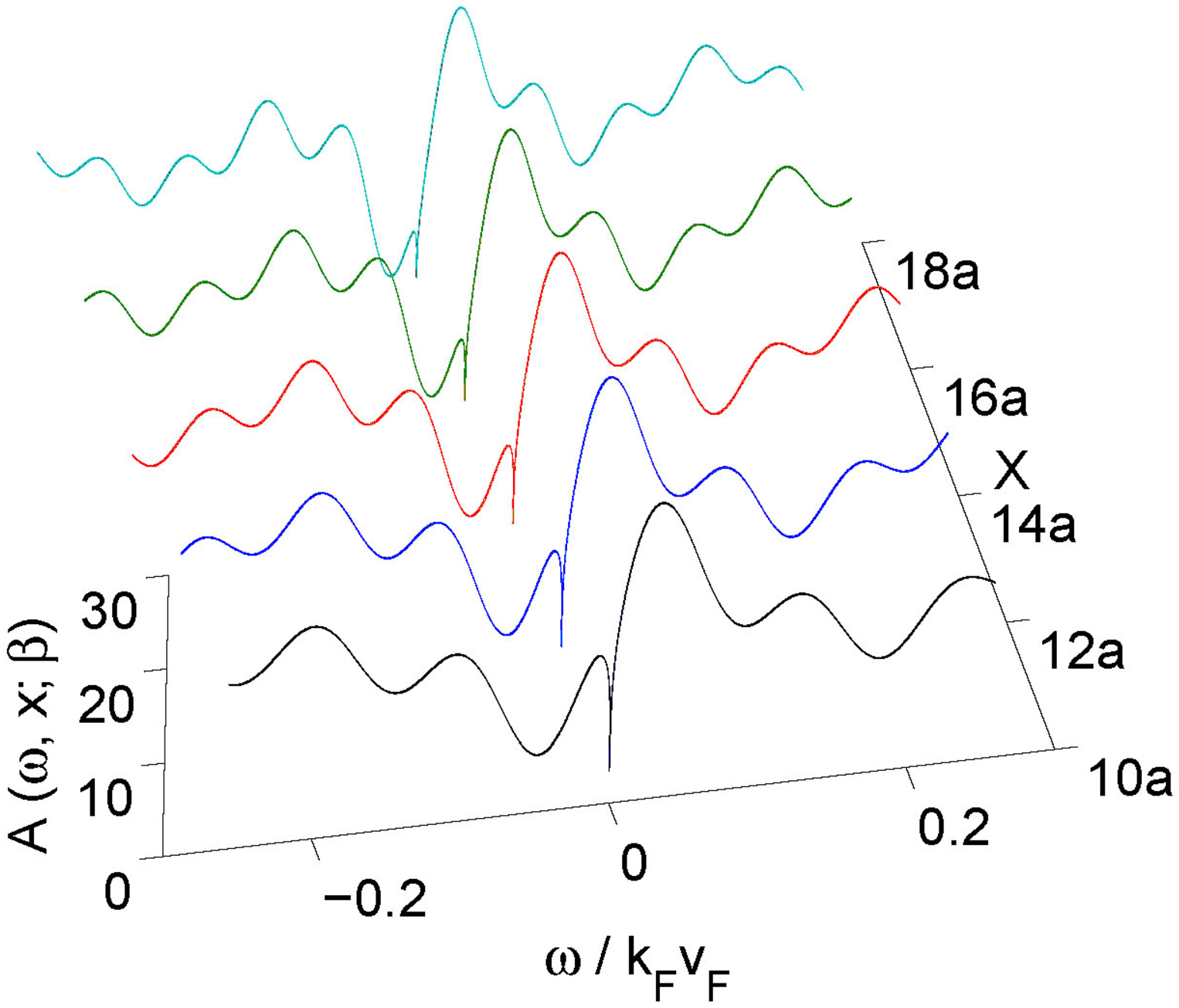}
\includegraphics[width=3.2in]{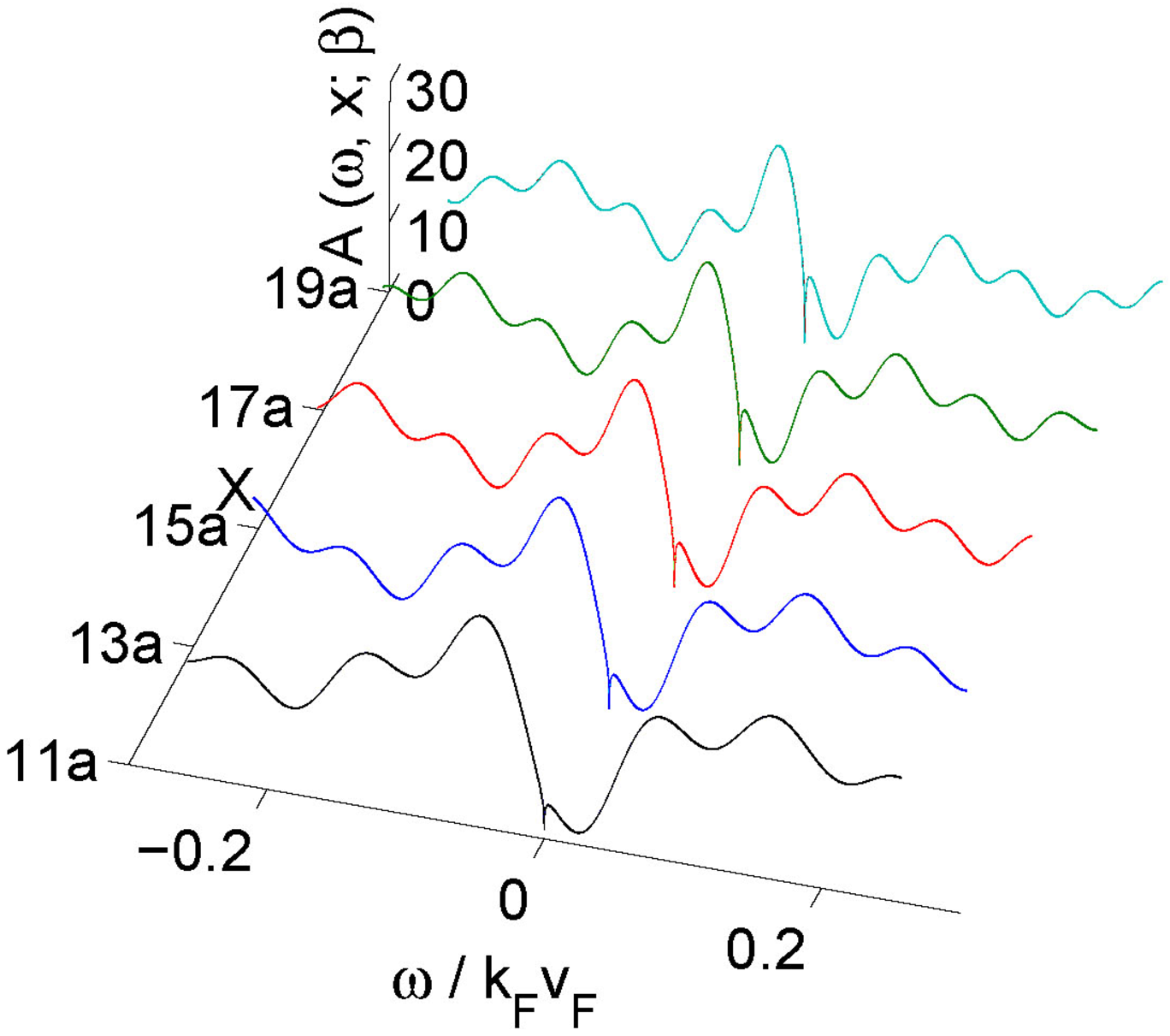}
\caption{[color online] $A(\omega, x; \beta)$ for $K_{c}=0.7$,
  $v_{c}/v_{F} \approx 1.43$, $v_{c}/v_{s} \approx 3$,
  $T/k_{F}v_{F}=2.6 \times 10^{-6}$, $n_{e}=0.97$, shown for even and
  odd sublattices.}
\label{Friedel_asymmetry}
\end{center}
\end{figure}
The shift depends upon the filling factor $n_e$ since,
for fixed $x$, $n_e$ determines the values of $k_F$ and $\gamma$ in
Eq.~(\ref{gamma}). The LSW is asymmetric in general, but for
particular fillings for which $2k_F n a + \gamma = \pi m \, (m \in
{\mathbb{Z}})$ it becomes a symmetric function of $\omega$. By
inspection of Eqs. (\ref{RealParts}) and (\ref{LDOSF}) one finds that
for a given interaction strength the asymmetry tends to zero very
close to the boundary ($2\omega x/v_{s} \ll 1$) and very far from the
boundary ($2 \omega x/v_{c} \gg 1$, ''bulk'' regime) and reaches a
maximum in the intermediate region. It is important to realize that
the shift of the periodic structure with respect to the Fermi level is
present also for the non-interacting case when the LSW takes the
simple form $A(\omega,x;\beta) \sim \cos(2k_Fx + \gamma -2\omega
x/v_{F})$. We conclude that the shift is a pure boundary effect, and
is due to the interference of the incoming and reflected electrons at
the boundary. In contrast, the dip of the spectral weight at the Fermi
level is an interaction effect.

In Fig.~\ref{Friedel_asymmetry} we show the energy and coordinate
dependence of the LSW for even and odd sublattices. The Friedel
oscillations on the scale of the lattice spacing $a$ are easily
visible as a flip of the asymmetry when going from one graph to the other.
The spin and charge modulations of the amplitude are also present over longer
wavelengths in real space, but are not clearly visible due to the relatively 
narrow coordinate range in Fig.~\ref{Friedel_asymmetry}.  Note that
the asymmetry with energy varies with the distance to the boundary
since the phase $2k_Fx + \gamma$ also varies with distance.
\begin{figure}[h!tb]
\begin{center}
\includegraphics[width=3.2in]{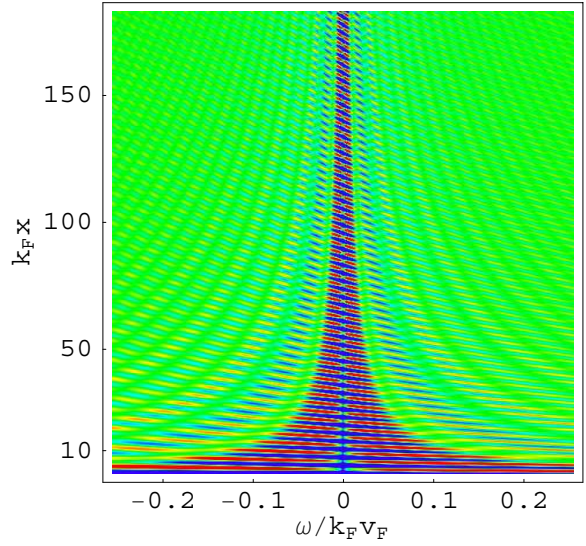}
\caption{[color online] $A(\omega, x; \beta)$ for $K_{c}=0.7$,
  $v_{c}/v_{F} \approx 1.43$, $v_{c}/v_{s} \approx 3$,
  $T/k_{F}v_{F}=2.6 \times 10^{-4}$, $n_{e}=0.97$.}
\label{boundary_2D}
\end{center}
\end{figure}

It is interesting to note the similarity of the typical asymmetric
line shapes in Figs.~\ref{asymmetry} and \ref{Friedel_asymmetry}
with that of a Fano resonance \cite{Fano} when $\omega$ is close to the Fermi
level. A Fano resonance is known to develop in the LSW for
non-interacting electrons when coupled e.g. to a magnetic impurity,
the effect being produced by the interference between resonating and
nonresonating electron paths through the impurity \cite{Ujsaghy}. As
we have seen, the asymmetry in the present case instead comes from the combined
effect of electron interactions (causing a dip in the LSW at the Fermi
level) and the reflection of electrons off the boundary (causing a
phase shifted oscillation in the LSW). As one expects the Fano line
shape to survive for {\em interacting} electrons coupled to a magnetic
impurity it is indeed satisfying to see this feature reproduced by the
open boundary for which the impurity gets traded at low temperatures.
\begin{figure}[h!tb]
\begin{center}
\includegraphics[width=3.0in]{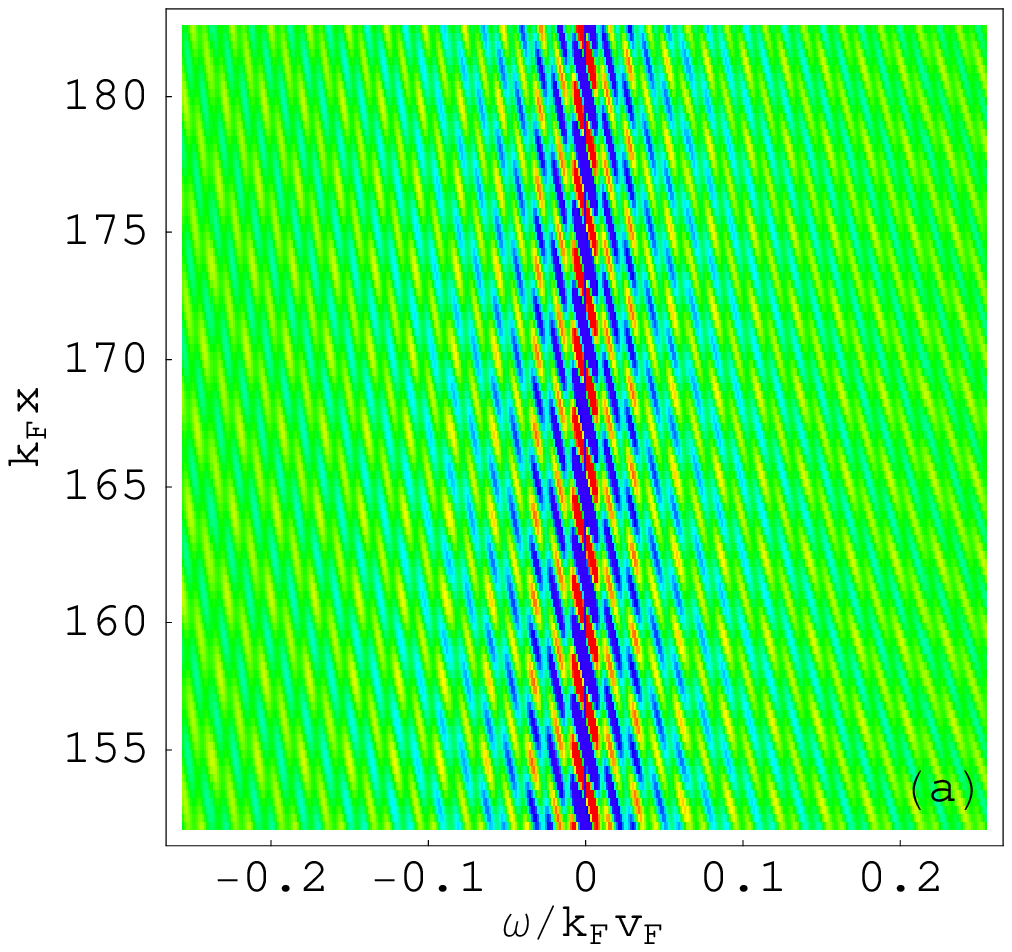}
\includegraphics[width=3.0in]{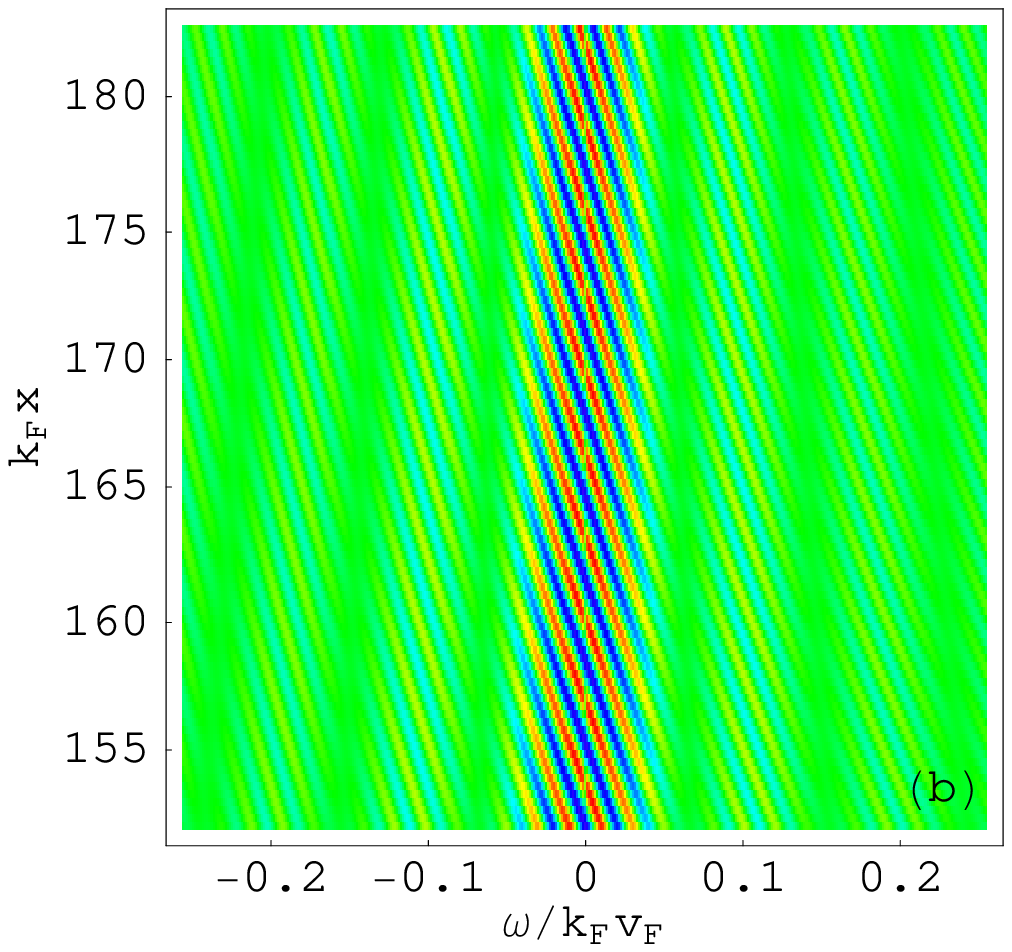}
\caption{[color online] $A(\omega, x; \beta)$ for  $T/k_{F}v_{F}=2.6
  \times 10^{-6}$, $n_{e}=0.97$, with $K_{c}=0.7$,  $v_{c}/v_{F} \approx 
  1.43$, $v_{c}/v_{s} \approx 3$ (a) and $K_{c}=0.9$, $v_{c}/v_{F} \approx
  1.11$, $v_{c}/v_{s} \approx 1.26$ (b). The color coding   is the
  same as in Fig.~\ref{boundary_2D}.}
\label{spin_charge_2D}
\end{center}
\end{figure}

\subsubsection{Spin-charge separation} 
\begin{figure}[!htb]
\begin{center}
\includegraphics[width=3.2in]{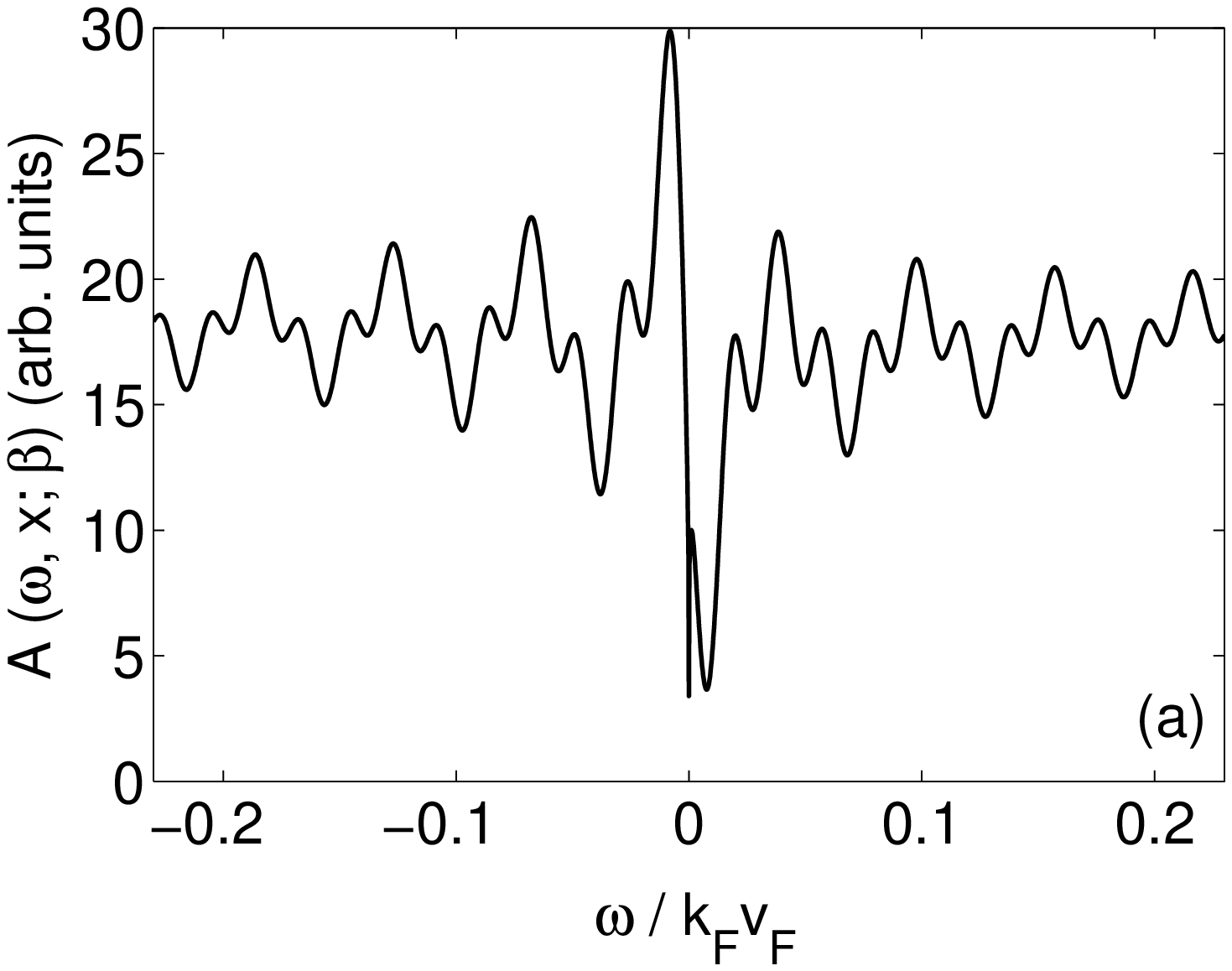}
\includegraphics[width=3.2in]{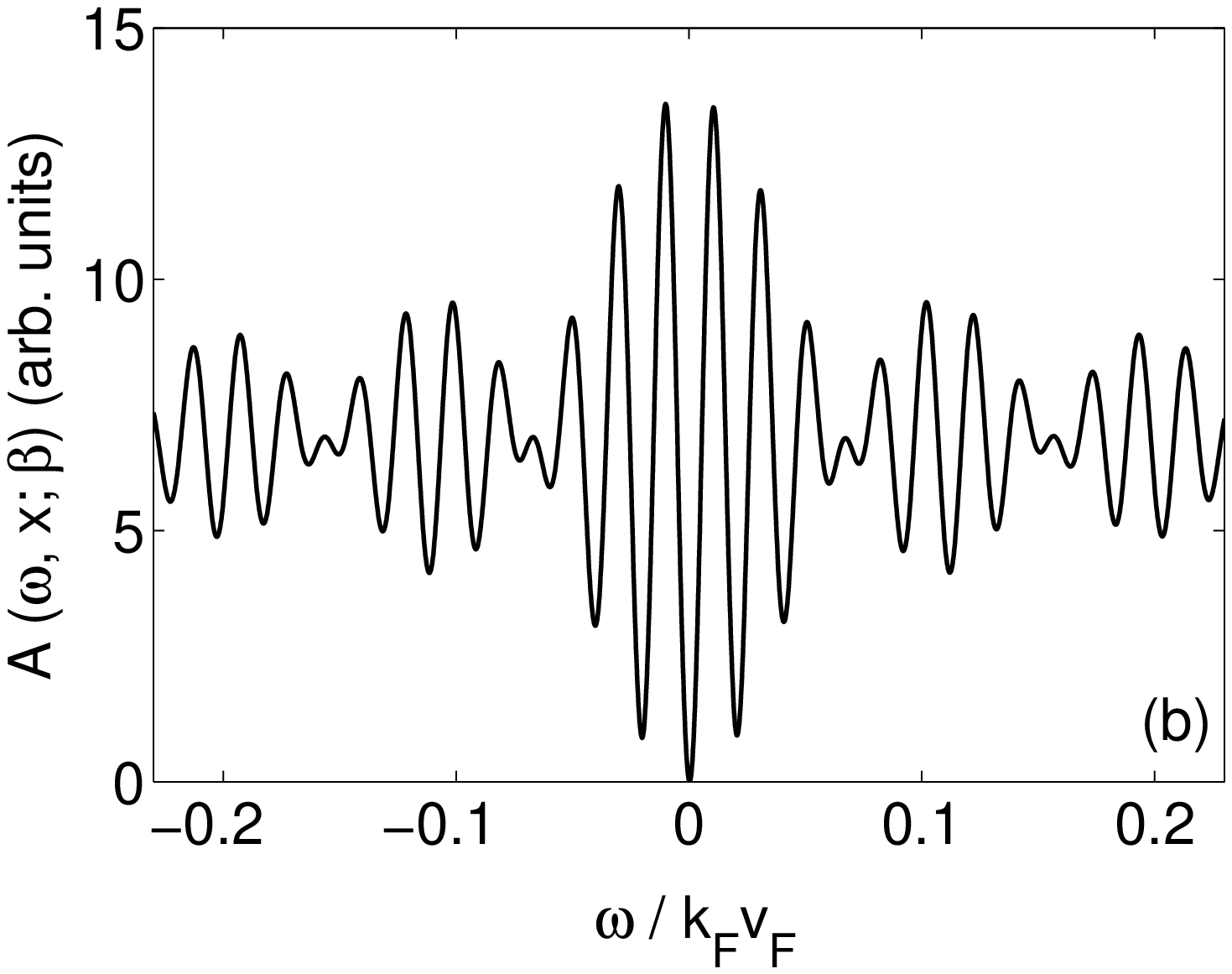}
\caption{$A(\omega, x; \beta)$ for $T/k_{F}v_{F}=2.6 \times 10^{-6}$,
  $n_{e}=0.97$, with 
  $x=50a$, $K_{c}=0.7$,  $v_{c}/v_{F} \approx
  1.43$, $v_{c}/v_{s}=\lambda_{c}/ \lambda_{s} \approx 3$ (a) and  
  $x=100a$, $K_{c}=0.9$, $v_{c}/v_{F} \approx
  1.11$, $v_{c}/v_{s}=\lambda_{c}/ \lambda_{s} \approx 1.26$ (b).}
\label{spin_charge}
\end{center}
\end{figure}

The proximity to an open boundary reveals a key property of interacting
electrons in one dimension $-$ spin-charge separation $-$ i.e. the
fact that the collective spin and charge excitations (induced e.g. by
inserting an extra electron into the system), propagate with different
speeds, and hence ''separate'' \cite{Voit}. The effect shows up in the LSW as a
characteristic peak structure at intermediate distances from the boundary.
Very far from the boundary ($\omega x_s  \gg 1$)
$A(r, \omega;\beta)$  is a monotone function scaling as
$\omega^{\alpha}$ near the Fermi level, with bulk
exponent $\alpha = (K_c + K_c^{-1})/4 - 1/2$ at low temperatures \cite{Voit}.
Extremely close to the boundary ($\omega
x_s  \ll 1$) $A(r, \omega;\beta)$ has a similar structure, but with an enhanced
suppression near the Fermi
level, $A(r, \omega; \beta) \sim \omega^{\alpha_B}$, with boundary exponent
$\alpha_B = (K_c^{-1} -1)/2$ at
low temperatures \cite{KaneFisher,FabrizioGogolin,EJM} (see
Fig.~\ref{boundary_2D}, with color coding in arbitrary units). As one
moves away from the immediate vicinity of the 
boundary an oscillation pattern emerges, which becomes most pronounced
when $\omega x_s  \sim {\cal O}(1)$.
This oscillatory feature (see Fig.~\ref{spin_charge_2D}) which is a
superposition of spin and charge waves is due to
the two types of branch points in Eqs. (\ref{GLL1}) and (\ref{GLL2})
which define the propagating collective spin and charge modes. The two panels
correspond to two different choices of $K_c$,
with two different values of the ratio $v_c/v_s$, leading to the two
different peak structures. By close
inspection of the graphs one can easily read off the corresponding
velocity ratios. To see how, let us make a
''cut'' of the two panels in Fig.~\ref{spin_charge_2D} at a distance
$x=100a$ from the boundary, yielding the panels in Fig.~\ref{spin_charge}. 

When $K_c = 0.7$ (panel (a)) the spin and charge velocities differ
significantly. For this case the short wavelength
($\lambda_{s}=\pi v_{s}/x $) spin oscillations are
modulated by long wavelength charge oscillations
($\lambda_{c}=\pi v_{c}/ x$). We note that there are
three spin oscillations per one charge oscillation which is in agreement with
the fact that in this case $\lambda_{c}/ \lambda_{s} = v_c/v_s \approx
3$ (see Fig.~\ref{spin_charge} (a)). When $K_c$ gets closer to unity
(non-interacting limit) the spin and charge velocities approach each other
and the spin-charge separation manifests itself as a beating pattern,
provided that $K_c$ is not identical to unity (see
Fig.~\ref{spin_charge} (b) where $K_c = 0.9$). The short wavelength
($\lambda=2\pi v_{s}v_{c}/ x(v_{s}+v_{c})$)
oscillations are now amplitude modulated by long wavelength oscillations
($\lambda'=2\pi v_{s}v_{c}/x(v_{c}-v_{s})
$). We see that there are about six short wavelength oscillations per
"bubble" of the long wavelength amplitude modulations, which is in
agreement with the fact $\lambda '/ \lambda \approx 12$ in this case,
corresponding to $\lambda_{c}/ \lambda_{s} = v_c/v_s \approx 1.26$.  

The spin-charge oscillations in the LSW are also present in real space
as discussed before \cite{Eggert}. The oscillations as a function of energy
might be easier to observe, however, since here there are no superimposed
Friedel oscillations. We also remark that the existence of the beating
pattern for values of $K_c$ close to unity was proposed in
Ref.~[\onlinecite{UssishkinGlazman}] as a diagnostic tool for
spin-charge separation in possible tunneling experiments of an LL
where a scanning probe microscopy tip would be used as an impurity.
\begin{figure}[!htb]
\begin{center}
\includegraphics[width=3.2in]{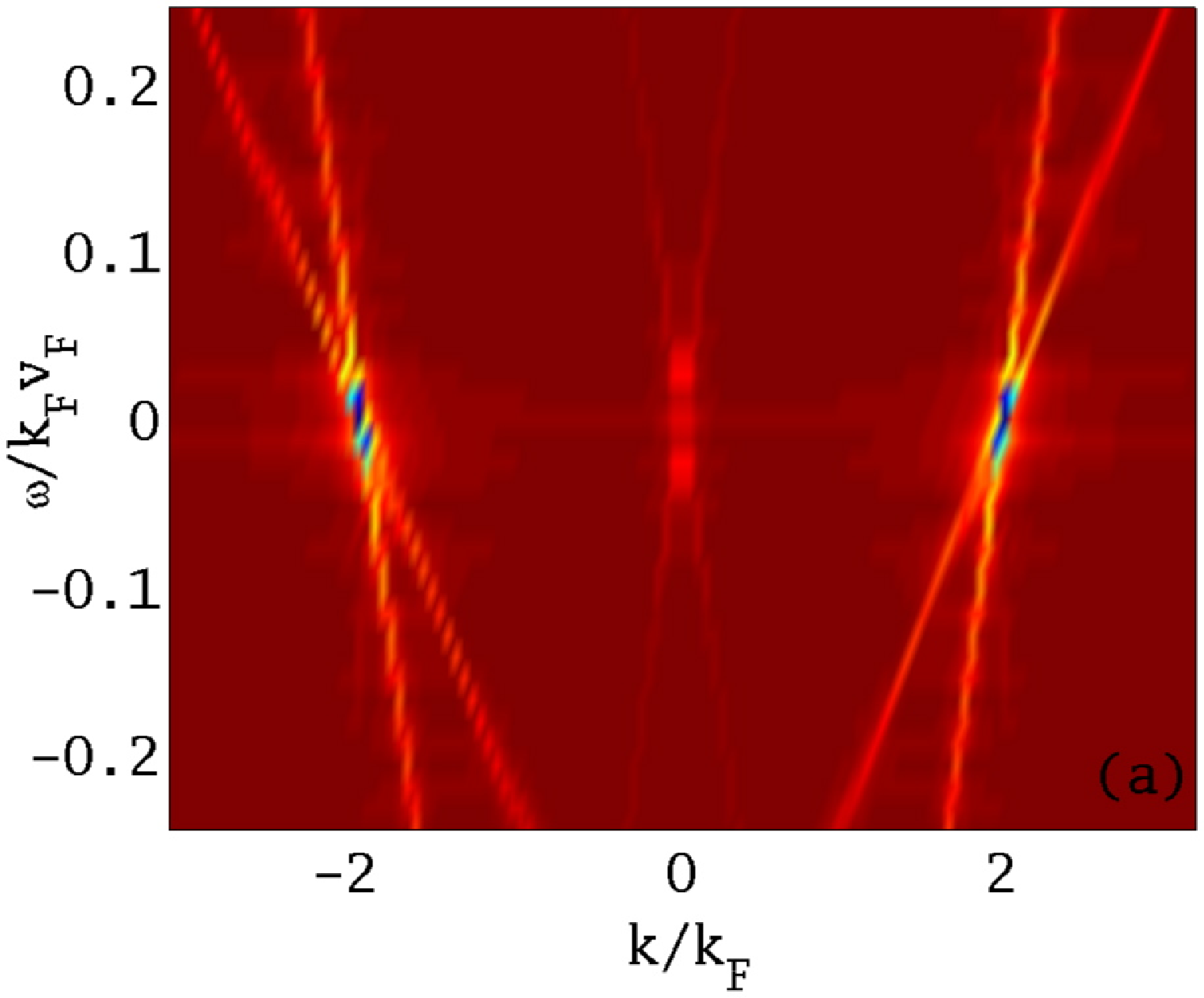}
\includegraphics[width=3.2in]{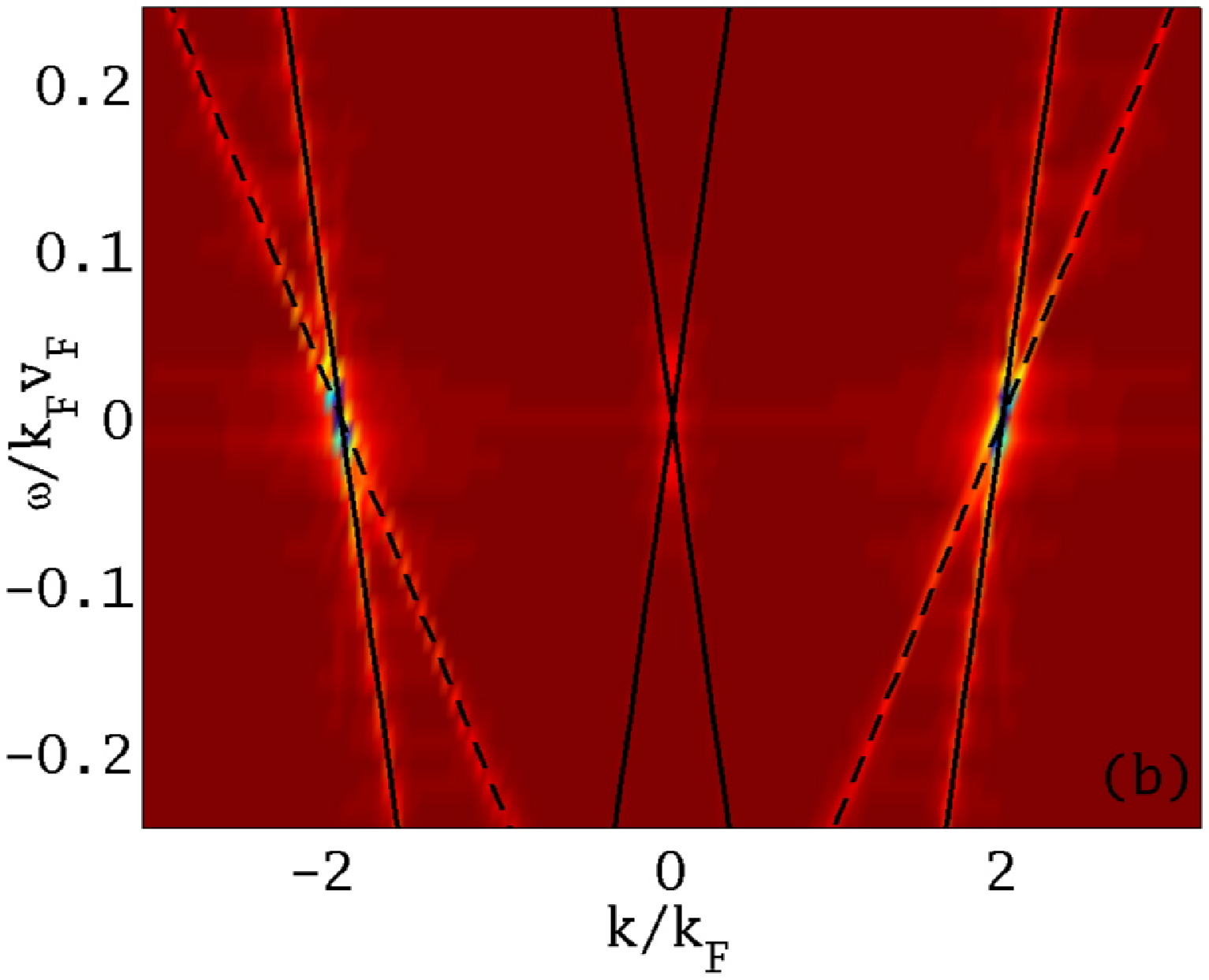}
\caption{Fourier transform of the LSW for $K_{c}=0.7$, $v_{c}/v_{F} \approx
  1.43$, $v_{c}/v_{s} \approx 3$, $T/k_{F}v_{F}=2.6 \times 10^{-4}$,
  $n_{e}=0.97$ (a) and 
  the same graph with charge (solid lines) and spin (dashed lines)
  wave dispersions superimposed (b).}
\label{FourierTrans}
\end{center}
\end{figure}

One can map the dispersion of the spin and charge waves by taking a
Fourier transform of the LSW (see Fig.~\ref{FourierTrans} (a)).  This Fourier 
transform should not be confused with the momentum or angle resolved
spectral weight, which is measured in photoemission experiments,
although it shows similar features of spin charge separation. The
dominant weights in the transform correspond to the $\omega (k)$
dependence of the excitations. The dispersion lines at $k=0$ come from the
non-oscillatory part of the LSW and represent the charge excitations,
since the non-oscillatory part contains only charge oscillations. This
feature at $k=0$ is also a clear indication of interaction effects and
disappears as $K_c \to 1$.  The dispersion lines at $k \ne 0$ come
from the Friedel terms, and contain spin and charge branches shifted
from $k=0$ by $\pm 2 k_{F}$. The mirror symmetry about $k=0$ reflects
the standing wave nature of the oscillations. In
Fig.~\ref{FourierTrans} (b) the dispersion relations $\omega(k)=\pm
kv_{c}/2$ at $k\approx 0$ and $\omega(k)=(\pm k-2k_{F})v_{c/s}/2$ at
$k\approx \pm 2k_{F}$ are plotted on top of the Fourier transform and
agree with the location of the maxima.

\subsubsection{Thermal effects} 
\begin{figure}[!htb]
\begin{center}
\includegraphics[width=3.2in]{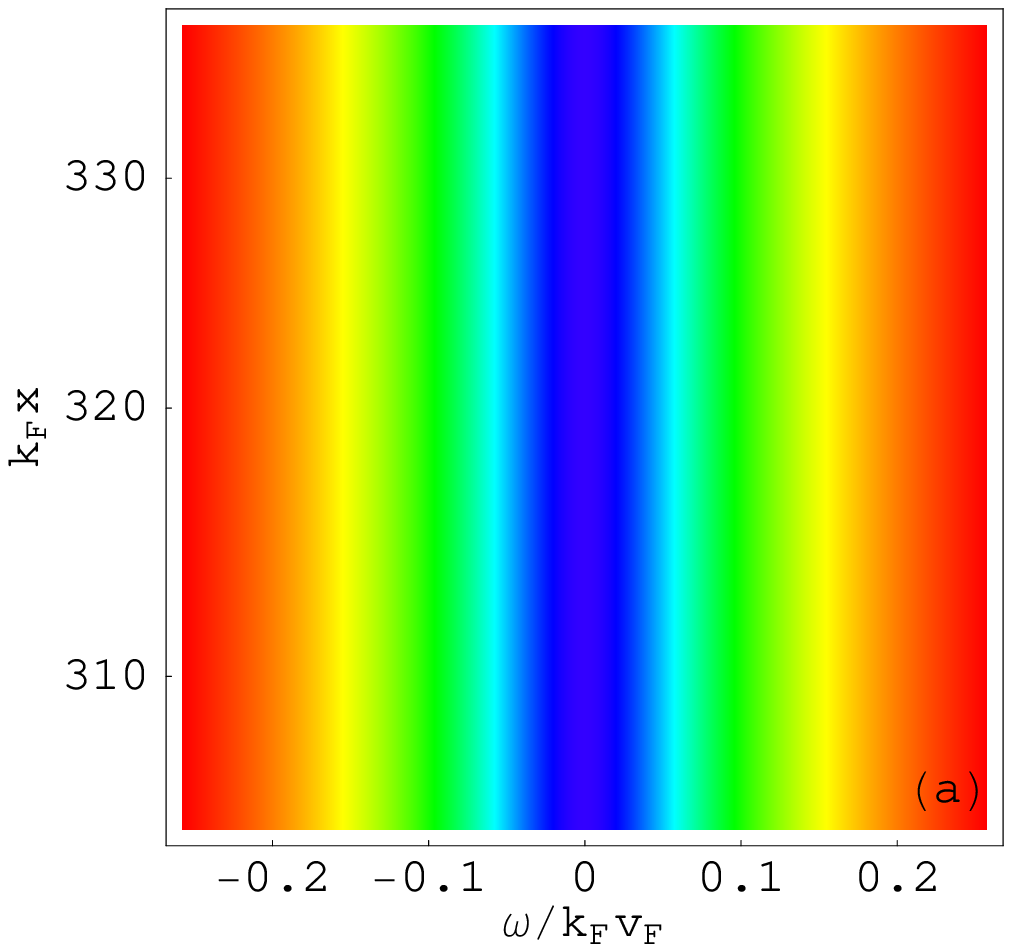}
\includegraphics[width=3.2in]{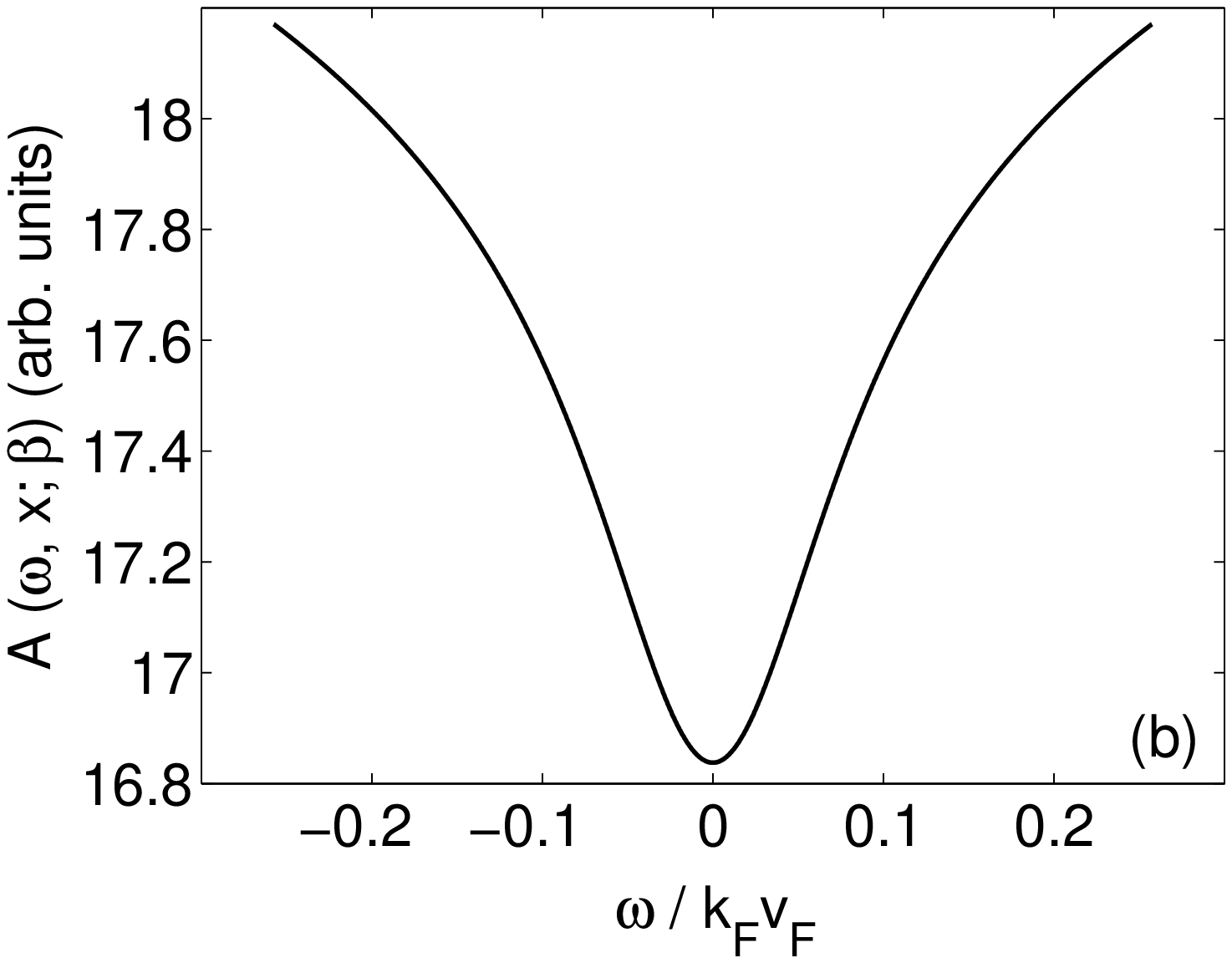}
\caption{[color online] $A(\omega, x; \beta)$ for $T/k_{F}v_{F}=2.6
  \times 10^{-2}$, $K_{c}=0.7$, $v_{c}/v_{F} \approx 1.43$, $v_{c}/v_{s}
  \approx 3$, $n_{e}=0.97$ (a) and cut at $x=200a$ (b).}
\label{NO_spin_charge_2D}
\end{center}
\end{figure}

On general grounds one expects that thermal effects become visible only for
energies $\omega \lesssim T$ and distances $x \gtrsim \lambda_T=v_s/T$.
Choosing e.g. $T\lesssim 10$K and $v_F \approx 10^5 \rm m/s$ (a
typical value for a quasi-1D 
organic metal for which LL theory should be
applicable \cite{Giamarchi}) this implies that the spin-charge peak
structure as 
seen in Fig.~\ref{spin_charge_2D} 
will remain intact for distances not too far from the edge
($2x/\beta v_{s} \ll 1$).
We should caution the reader that the energy range for which the 
LL theory is applicable to a specific experiment may sometimes be
smaller than the range depicted in Fig.~\ref{spin_charge_2D}
(cf. our discussion in Sec. I). 
For $2x/\beta v_{c} \gtrsim 1$ the spin and charge waves loose their
coherence before reaching the edge and therefore spin-charge separation
is destroyed (see Fig.~\ref{NO_spin_charge_2D}(a)). 
\begin{figure}[!htb]
\begin{center}
\includegraphics[width=3.2in]{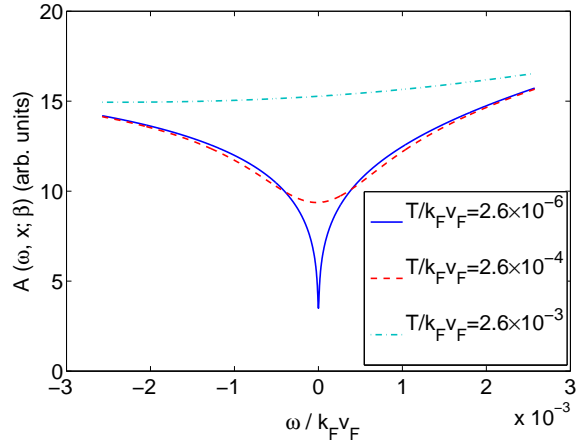}
\caption{[color online] $A(\omega, x; \beta)$ for $K_{c}=0.7$, $v_{c}/v_{F} \approx
  1.43$, $v_{c}/v_{s} \approx 3$, $x=10a$, $n_{e}=0.97$ at different
temperatures.}
\label{temperature}
\end{center}
\end{figure}

A more interesting issue is the
fate of the asymptotic scaling behavior of the LSW near the Fermi level as the
temperature increases. In
earlier work it was found that the ''uniform part'' of the LSW
(corresponding to the first term in (\ref{LDOSF})
crosses over to $\omega^2$ scaling in both boundary ($2 \omega x/v_{s} \ll
1$) and bulk regimes ($2 \omega x/v_{c} \gg 1$)
when $\omega \beta < 1$ \cite{MEJ}. The effect was found to originate in the
exponential damping of the
density correlations for unequal times due to thermal fluctuations. By
performing an expansion of the {\em
full} LSW in Eq. (\ref{LDOSF}) with the small parameter $\omega \beta$
we find that the power law is now modified to 
\begin{equation}  \label{ThermalScaling}
A(\omega, x; \beta) \sim A + B\omega + C\omega^2, \ \ \ \ \omega \beta \ll 1
\end{equation}
where $A, B$ and $C$ depend on the temperature and the distance from
the boundary. We depict the crossover
from boundary scaling $\sim \omega^{(K_c^{-1} -1)/2}$  for $\omega
\beta \gg 1$ to thermal scaling $\sim A + B\omega +
C\omega^2$ for $\omega \beta \ll 1$ in Fig.~\ref{temperature}. Note that one
is able to observe this  crossover for $2x/\beta v_{s} \ll 1$. For
$2x/\beta v_{c} \gtrsim 1$ the
boundary scaling is completely washed out. Since in this regime
$G_{LL}(t,x,-x;\beta)$ in Eq.~(\ref{GLL2}) is exponentially suppressed
compared to $G_{LL}(t,x,x;\beta)$ in (\ref{GLL1}), it follows that
$A(\omega, x; \beta) \sim A + C\omega^2$, and the line shape becomes
symmetric (see Fig.~\ref{NO_spin_charge_2D}b).

In this context it is important to point out that our finite-temperature results
are strictly valid only for the case where an {\em edge} is modeled by an open
boundary. An {\em impurity}, on the other hand, is faithfully represented by an
open boundary condition only in the zero-temperature limit
\cite{KaneFisher,FurusakiNagaosa1}. For finite temperatures a new energy scale
$T_0$ appears, characterizing the crossover from weak to strong coupling
(''open boundary fixed point''), and depending on the impurity strength $V_0$.
A simple RG estimate shows that $T_0$ scales with $V_0$ as
$T_0 \sim V_0^{2/(1-K_{c})}$, with an overall scale factor that depends on the
details of the regularization procedure \cite{FurusakiNagaosa1}. For finite $T$ with $T\ll T_0$,
an open boundary representation of the impurity is still expected to capture the
essential physics. An interesting, albeit technically challenging project would be
to redo the calculations in this section for an impurity in the presence of the 
operators that appear away from the $T=0$ open boundary fixed point, allowing for a
complete picture of impurity thermal effects.

\subsection{Properties of the Local Tunneling Conductance}

\begin{figure}[!htb]
\begin{center}
\includegraphics[width=3.2in]{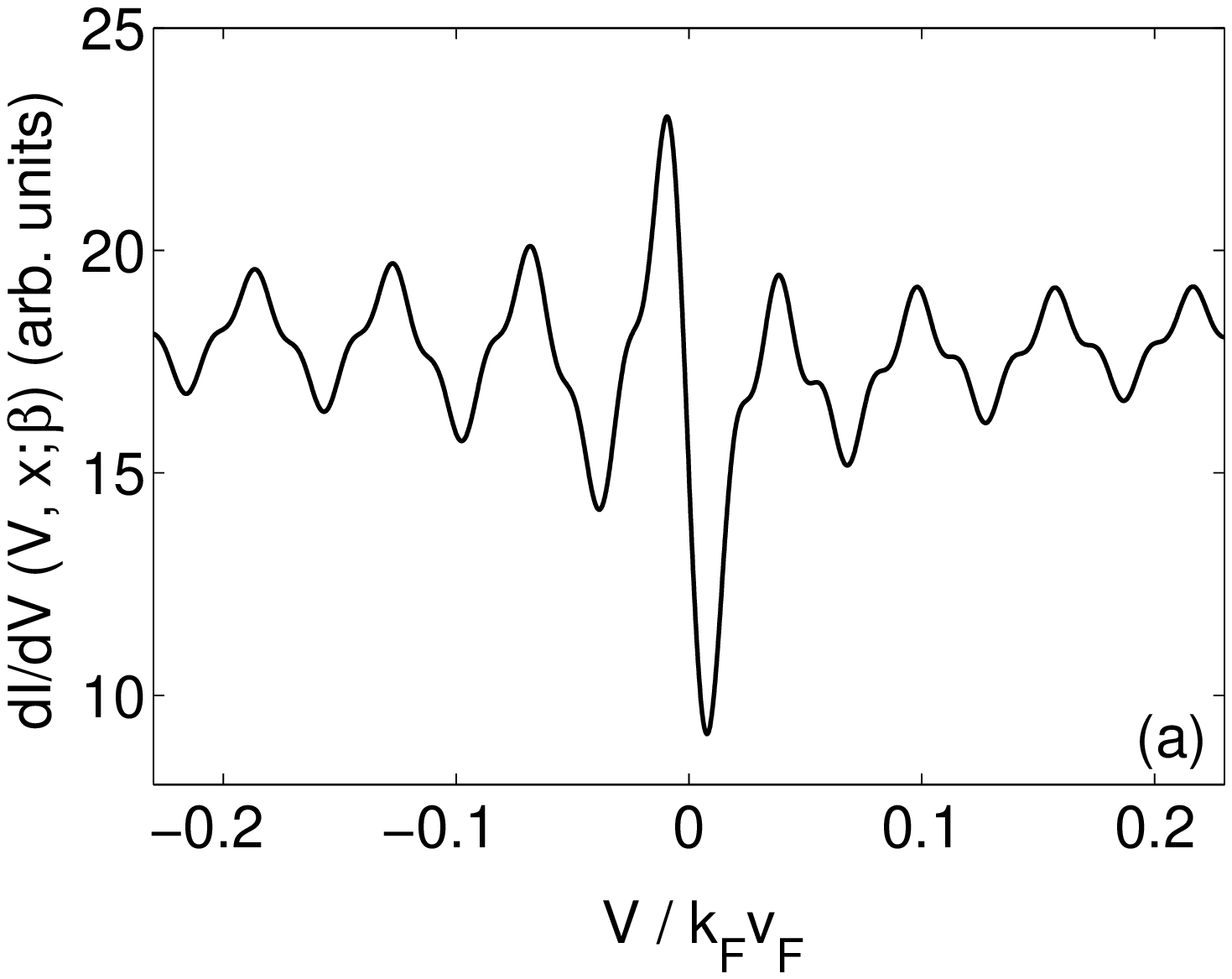}
\includegraphics[width=3.2in]{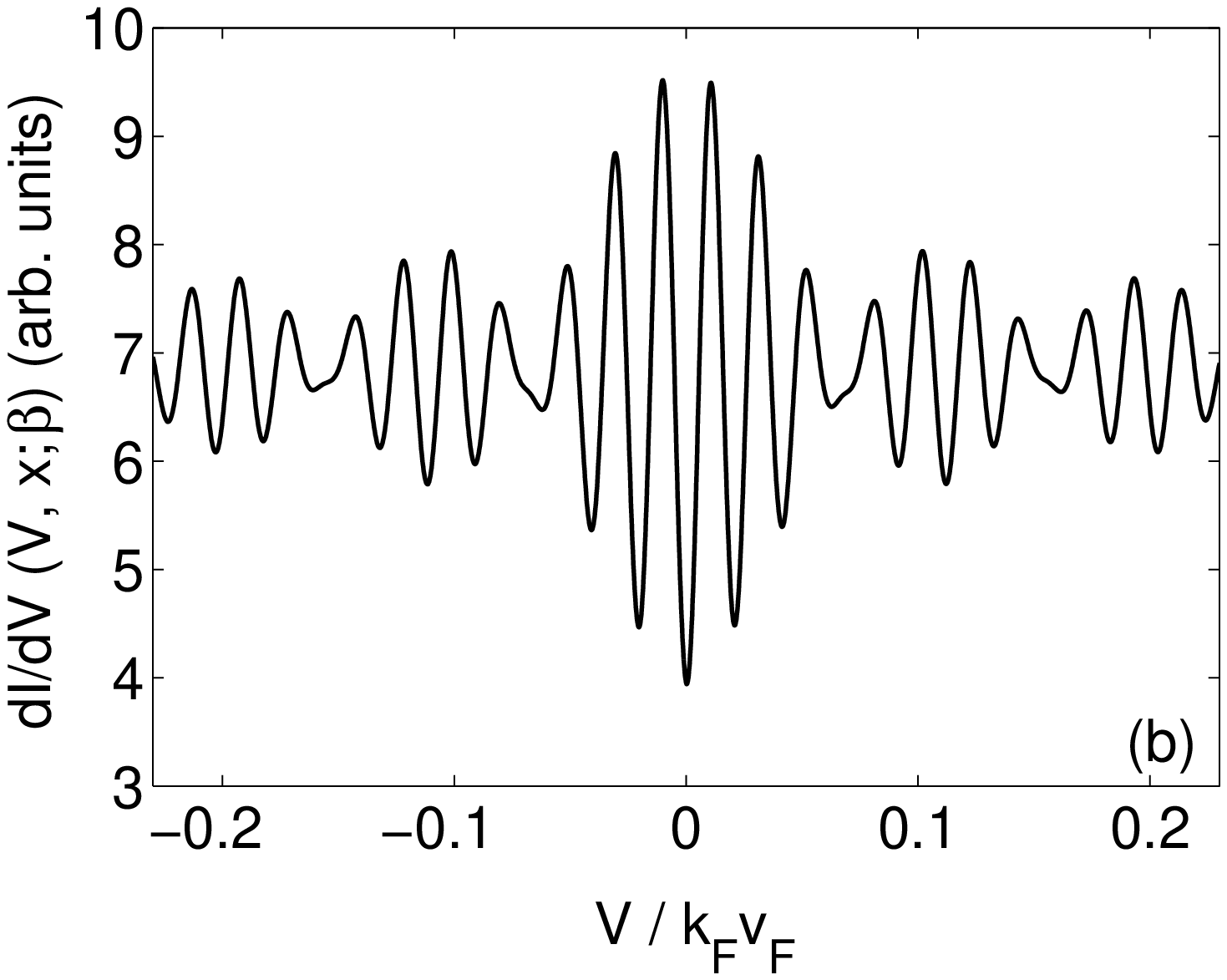}
\caption{$dI/dV(V, x; \beta)$ for $T/k_{F}v_{F}=2.6 \times 10^{-3}$,
  $n_{e}=0.97$, with
  $x=50a$, $K_{c}=0.7$,  $v_{c}/v_{F} \approx
  1.43$, $v_{c}/v_{s}=\lambda_{c}/ \lambda_{s} \approx 3$ (a) and  
  $x=100a$, $K_{c}=0.9$, $v_{c}/v_{F} \approx
  1.11$, $v_{c}/v_{s}=\lambda_{c}/ \lambda_{s} \approx 1.26$ (b).} 
\label{dIdV}
\end{center}
\end{figure}

The \emph{local differential tunneling conductance}, defined in 
Eq.~(\ref{LDOST}), exhibits the very 
same features as the LSW. The only difference is that the fine
structure of the  differential conductance is thermally smeared via
the temperature dependence of the Fermi-Dirac distribution.  This can
clearly be seen by examining the energy dependence at some fixed
distance from the boundary, as done in Fig.~\ref{dIdV}, and then
comparing to the corresponding graphs for the LSW in Fig.~\ref{spin_charge}.
As the smearing occurs on
the scale of $T$, spin-charge separation is wiped out for temperatures
$T \gtrsim \lambda_{s}$.   The distance $x$ from the boundary determines the 
wavelength of the oscillations in energy space, so in most cases it should be 
possible to find a range for $x$ which shows many waves in the energy
interval where LL theory applies that are not washed out by
temperature, i.e. $1/\Delta \ll x/v_c \ll 1/T$, where $\Delta$ 
is the bandwidth.  Note that the inequality above coincides with the criterion 
for observing spin-charge separation in the LSW, discussed in the
previous subsection.

In wires of finite length $l$ the energy levels become discrete and
are given by a characteristic direct product of two spectra 
with uniform spacing $\pi v_c/l$ and $\pi v_s/l$ \cite{MEJ,AnfusoEggert}. The
standing waves of the individual levels show the corresponding intereference 
of spin and charge excitations \cite{AnfusoEggert}. If the temperature
becomes comparable to the level spacing, a continous pattern of
Friedel oscillations as shown above emerges.  However, in order to see
the predicted behavior at all, the wire must always be long enough so that many
levels lie within the energy interval in which the LL theory applies,
i.e.~$\pi v_c/l \ll \Delta$. For metallic materials this translates into lengths
of typically 100nm or more.

When applying the results to SWCNTs similar features can be expected
as has already been partially confirmed experimentally \cite{Lee}.
The dominant effect is the complicated interference pattern
of the Bloch waves, which strongly depends on the geometry of the
boundary condition and the chirality of the tubes \cite{mele}.
Nonetheless, an enhanced velocity, a suppressed spectral weight and a
characteristic power law of decaying oscillations are clear signatures
of interaction effects which all have been seen in experiments
\cite{Lee}. A more complete theoretical analysis would also have to
include the effects of backscattering, band structure, longer-range
interactions and the mixing of the channels near the boundary which we
defer to a future publication.

\section{Summary}

We have derived the full finite-temperature LSW for a 
Luttinger liquid with an edge or impurity (modeled as an open boundary
condition), relevant to high-precision STM measurements. We have also
generalized our approach to the ''two-channel'' case that describes 
SWCNTs in the Luttinger liquid regime, which is qualitatively similar to the 
single-channel case.  

The LSW (determining the local differential tunneling conductance in
STM measurements) exhibits a very rich structure
as a function of temperature, distance from the impurity, and the strength
of the electron interaction. Depending on the choice of
parameters one is able to see \emph{asymmetric Fano-like line shapes} and
\emph{spin and charge oscillations}. The Fano-like asymmetries are
caused by an interplay of boundary and interaction effects, and, as we
have shown, are closely linked to the Friedel oscillations in real
space. Spin and charge oscillations appear due to the interference of
propagating spin and charge waves reflected from the boundary.

We have discussed how to consistently determine the key parameters 
of a Luttinger liquid (the interaction parameter $K_c$, and the spin
and charge velocities) from experimental measurements of the tunneling
conductance. We have also extensively discussed various thermal
effects with focus on their influence on the Fano-like asymmetries and
spin-charge oscillations in the LSW as a function of energy. The
thermal suppression of the coherence of spin
and charge waves makes it hard to detect interaction effects in
the LSW and even more so in the local differential tunneling conductance (where
the finite-temperature LSW gets weighted by 
the Fermi-Dirac distribution): For temperatures $T \gtrsim v_s/x$
(where $v_s$ is the speed of the spin excitations, and $x$ is the
distance to the edge or the impurity) the characteristic power law
behavior near the Fermi level is completely washed out and replaced by
an interaction-independent analytic scaling. For these temperatures
spin-charge separation also becomes virtually impossible to detect.  

In conclusion, our results provide guidelines for identifying and interpreting
signals of electron correlations in STM data on SWCNTs and other
one-dimensional systems, and as such should be useful in the search
for realizations of Luttinger liquid physics. 

\begin{acknowledgments}
It is a pleasure to thank Reinhold Egger for valuable input at the
early stages of this project. 
We are also indebted to Andrew Green for helpful discussions.
 Financial support from the Swedish Research Council is acknowledged.
\end{acknowledgments}



%
%

\begin{thebibliography}{99}
\bibitem{Giamarchi}
T.~Giamarchi, {\em Quantum Physics in One Dimension} (Oxford University Press, Oxford,
 2004).

\bibitem{Voit}
J.~Voit, Rep. Prog. Phys. {\bf 58}, 977 (1995).

\bibitem{KaneFisher}
C.~L.~Kane and M.~P.~A.~Fisher, Phys. Rev. B {\bf 46}, 15233 (1992).

\bibitem{FurusakiNagaosa1}
A.~Furusaki and N.~Nagaosa, Phys. Rev. B {\bf 47}, 4631 (1993).

\bibitem{FurusakiNagaosa2} 
A.~Furusaki and N.~Nagaosa, Phys. Rev. Lett. {\bf 72}, 892 (1994).

\bibitem{EggertAffleck}
S.~Eggert and I.~Affleck, Phys. Rev. B {\bf 46}, 10866 (1992).

\bibitem{FabrizioGogolin}
M.~Fabrizio and A.O.~Gogolin, Phys. Rev. B {\bf 51}, 17827 (1995). 

\bibitem{EJM}
S.~Eggert, H.~Johannesson, and A.~Mattsson, Phys. Rev. Lett. {\bf 76}, 1505 (1996).

\bibitem{MEJ}
A.~E.~Mattsson, S.~Eggert, and H.~Johannesson, Phys. Rev. B {\bf 56}, 15615 (1997).

\bibitem{VoitWangGrioni}
J.~Voit, Y.~Wang, and M.~Grioni, Phys. Rev. B {\bf 61}, 7930 (2000).

\bibitem{WangVoitPu}
Y.~Wang, J.~Voit, and F.-C.~Pu, Phys. Rev. B {\bf 54}, 8491 (1996).

\bibitem{AnfusoEggert}
F.~Anfuso and S.~Eggert, Phys. Rev. B {\bf 68}, 241301(R) (2003).

\bibitem{Schonhammer1}
K.~Sch\"onhammer, V.~Meden, W.~Metzner, U. Schollw\"ock, and O. Gunnarsson,
Phys. Rev. B {\bf 61}, 4393 (2000).

\bibitem{Meden1}
V.~Meden, W.~Metzner, U.~Schollw\"ock, O.~Schneider, T.~Stauber, and
K.~Sch\"onhammer, Eur. Phys. J. B {\bf 16}, 631 (2000).

\bibitem{Bockrath}
M.~Bockrath, D.~H.~Cobden, H.~Lu, A.~G.~Rinzler, R.~E.~Smalley, L.~Balents, and
P.~L.~McEuen,
Nature {\bf 397}, 598 (1999).

\bibitem{Yacoby}
A.~Yacoby, H.~L.~Stormer, N.~S.~Wingreen, L.~N.~Pfeiffer, K.~W.~Baldwin, and
K.~W.~West,
Phys. Rev. Lett. {\bf 77}, 4612 (1996).
\bibitem{Segovia}
P.~Segovia, D.~Purdie, M.~Hengsberger, and Y.~Baer, Nature {\bf 402}, 504 (1999).
\bibitem{Eggert}
S.~Eggert, Phys. Rev. Lett. {\bf 84}, 4413 (2000).

\bibitem{Kivelson}
S.~A.~Kivelson, I.~P.~Bindloss, E.~Fradkin, V.~Oganesyan, J.~M.~Tranquada, A.~Kapitulnik, and
C.~Howald, 
Rev. Mod. Phys. {\bf 75}, 1201 (2003).

\bibitem{CorrectionFootnote}
Note that the boundary exponent in Ref.~[\onlinecite{EJM}] is expressed via
a non-standard parameterization of the LL interaction. 

\bibitem{Yao}
Z.~Yao, H.~W.~Ch.~Postma, L.~Balents, and C.~Dekker, Nature {\bf 402},
273 (1999). 

\bibitem{Auslaender}
O.~M.~Auslaender, A.~Yacoby, R.~de Picciotto, K.~W.~Baldwin,
L.~N.~Pfeiffer, and K.~W.~West, Science {\bf 295}, 825 (2002).

\bibitem{Zwick}
F.~Zwick, S.~Brown, G.~Margaritondo, C.~Merlic, M.~Onellion, J.~Voit, and M.~Grioni, 
Phys. Rev. Lett. {\bf 79}, 3982 (1997).

\bibitem{Gunnarsson}
For a review, see O.~Gunnarsson, K.~Sch\"onhammer, J.~W.~Allen, K.~Karlsson, and
O.~Jepsen,
J. Electr. Spectr. {\bf 117}, 1 (2001). 

\bibitem{SPM}
An alternative setup which avoids this problem is to use the scanning tip to create
a local potential with a scanning probe microscopy tip which scatters
electrons, and then measure how the tunneling into the system from a fixed tunnel
junction depends on the distance from the tip position as suggested by
 Ref. [\onlinecite{UssishkinGlazman}].

\bibitem{Venema}
L.~C.~Venema, J.~W.~G.~Wildoer, J.~W.~Janssen, S.~J.~Tans, H.~L.~J.~T.~Tuinstra, L.~P.~Kouwenhoven, and C.~Dekker, 
Science {\bf 283}, 52 (1998).

\bibitem{Lemay}
S. G. Lemay, J.~W.~Janssen, M.~van den Hout, M.~Mooij, M.~J.~Bronikowski, P.~A.~Willis, R.~E.~Smalley, 
L.~P.~Kouwenhoven, and C.~Dekker,
Nature {\bf 412}, 617 (2001).

\bibitem{LeRoy}
B.~J. LeRoy, S.~G. Lemay, J. Kong, and C. Dekker, 
Appl. Phys. Lett. {\bf 84}, 4280 (2004)

\bibitem{Lee1}
J.~Lee, H.~Kim, S.-J.~Kahng, G.~Kim, Y.-W.~Son, J.~Ihm, H.~Kato, Z.~W.~Wang, T.~Okazaki, H.~Shinohara,
and Y.~Kuk,
Nature (London) {\bf 415}, 1005 (2002).

\bibitem{Lee}
J.~Lee, S.~Eggert, H.~Kim, S.~J.~Kahng, H. Shinohara, and Y. Kuk,
Phys. Rev. Lett. {\bf 93}, 166403 (2004).

\bibitem{Penc}
K.~Penc, K.~Hallberg, F.~Mila, and H.~Shiba, Phys. Rev. Lett. {\bf 77}, 1390 (1996);
Phys. Rev. B {\bf 55}, 15475 81997).

\bibitem{Meden2}
V.~Meden, W.~Metzner, U.~Schollw\"ock, and K. Sch\"onhammer,
Phys. Rev. B {\bf 65}, 045318 (2002).

\bibitem{EggerGogolinEPJ}
R.~Egger and A.~O.~Gogolin, Eur. Phys. J. B {\bf 3}, 281 (1998).

\bibitem{Fano} U.~Fano, Nuovo Cimento {\bf 12}, 156 (1935);
  Phys. Rev. {\bf 124}, 1866 (1961).

\bibitem{Ujsaghy}
O.~{\'U}js{\'a}ghy, J.~Kroha, L.~Szunyogh, and A.~Zawadowski,
Phys. Rev. Lett. {\bf 85}, 2557 (2000). 

\bibitem{Rickayzen} G.~Rickayzen, {\em Green's Functions and Condensed Matter} 
(Academic Press, London, 1980).

\bibitem{Tersoff}
J.~Tersoff and D.~R.~Hamann, Phys. Rev. B {\bf 31}, 805 (1985).

\bibitem{GNT}
A.~O.~Gogolin, A.~A.~Nersesyan, and A.~M.~Tsvelik, {\em Bosonization and Strongly
Correlated Systems}
(Cambridge University Press, Cambridge, 1998).

\bibitem{Schulz}
H.~J.~Schulz, Phys. Rev. Lett. {\bf 71}, 1864 (1993).

\bibitem{DiFransesco}
P.~Di~Fransesco, P.~Mathieu, and D.~Senechal, {\em Conformal Field Theory} (Springer
Verlag, New York, 1997).

\bibitem{Mahan}
G.~Mahan, {\em Many-Particle Physics} (Plenum, New York, 1981).

\bibitem{EggerGogolinPRL}
R.~Egger and A.~O.~Gogolin, Phys. Rev. Lett. {\bf 79}, 5082 (1997).

\bibitem{BalentsFisher}
L.~Balents and M.~P.~A.~Fisher, Phys. Rev. B {\bf 55}, R11973 (1997).

\bibitem{mele} C.L.~Kane and E.J.~Mele, Phys. Rev. B {\bf 59}, R12759
  (1999).

\bibitem{UssishkinGlazman}
I.~Ussishkin and L.~I.~Glazman, Phys. Rev. Lett. {\bf 93}, 196403 (2004).









\end{thebibliography}
\end{document}